\pdfoutput=1    
\documentclass[useAMS]{mn2e}
\usepackage{graphicx}  
\usepackage{mnras_cite}

\usepackage{dcolumn}  
\usepackage{bm}       
\usepackage{amssymb,amsmath}
\usepackage{color}

\newcommand{\angstrom}{\mbox{\normalfont\AA}}

\newcommand{\bea}{\begin{eqnarray}}
\newcommand{\eea}{\end{eqnarray}}

\newcommand{\beq}{\begin{equation}}
\newcommand{\eeq}{\end{equation}}

\newcommand{\beqa}{\begin{eqnarray}}
\newcommand{\eeqa}{\end{eqnarray}}
\newcommand{\camb}{{\tt CAMB$\_$sources}}

\def\la{\mathrel{\mathpalette\fun <}}
 \def\ga{\mathrel{\mathpalette\fun >}}
\def\fun#1#2{\lower3.6pt\vbox{\baselineskip0pt\lineskip.9pt
\ialign{$\mathsurround=0pt#1\hfil##\hfil$\crcr#2\crcr\sim\crcr}}}

\begin{document}

\title[RSD from photometric populations]
{Redshift-space distortions from the cross-correlation of  photometric populations}
\author[Asorey et al.] 
{\parbox{\textwidth}{Jacobo Asorey$^{1,2}$, Martin Crocce$^1$, Enrique
    Gazta\~naga$^1$
}	\vspace{0.35cm}\\
$^1$Institut de Ci\`encies de l'Espai (ICE, IEEC/CSIC),  E-08193 Bellaterra (Barcelona), Spain \\
$^2$Department of Physics, University of Illinois, 1110 W. Green St.,Urbana, IL 61801, USA.\\
}

\date{\today}
\volume{445}\pagerange{2825--2835} \pubyear{2014}
\maketitle

\begin{abstract} \\

Several papers have recently highlighted the
possibility of measuring redshift space distortions from 
angular auto-correlations of galaxies
in photometric redshift bins. In this work we extend this
idea to include as observables the cross-correlations between redshift bins, as
an additional way of measuring  radial information. 
We show that this extra information allows to reduce the
recovered error in the growth rate index $\gamma$ by a factor of $\sim
2$. Although the final error in $\gamma$ depends on the bias and the
mean photometric accuracy of the galaxy sample, the improvement from adding
cross-correlations is robust in different settings.
Another factor of $2-3$ improvement in the determination of 
$\gamma$ can be achieved
by considering two galaxy populations over the same photometric
sky area but with different biases. This
additional gain is shown to be much larger than the one from
the same populations when observed over different areas of the sky
(with twice the combined area). The total improvement of $\sim 5$ implies
that a photometric survey such as the Dark Energy Survey should be able to recover 
$\gamma$ at the $5-10\%$ from the angular clustering in linear scales
of two different tracers. It can also constrain the evolution
of $f(z)\times\sigma_8(z)$ in few bins beyond $z\sim 0.8-0.9$ at the
$10-15\%$ level per-bin,
compatible with recent constrains from lower-$z$ spectroscopic
surveys. We also show how further improvement can be achieved by
reducing the photometric redshift error.

\end{abstract}


\section{Introduction}\label{sec:introduction}

Our understanding of the local Universe and the way it evolved from
small perturbations has been reshaped over the past decades with the
successful completion of vast
observational campaigns for CMB fluctuations, large scale structure
and SNIa distances. Yet several still open issues arose from
these studies, the most important of which is probably the late-time
accelerated expansion of the Universe. 

Hence many other cosmic surveys are ongoing or planned for the near future to
address these questions with a set of precision measurements never achieved
before. Several photometric surveys stand out among these, such as the
Dark Energy Survey (DES)\footnote{\tt www.darkenergysurvey.org}, the Panoramic Survey
Telescope and Rapid Response System
(PanStarrs)\footnote{\tt pan-starrs.ifa.hawaii.edu}, the Physics of the
Accelerating Universe survey (PAU)\footnote{\tt www.pausurvey.org},
and the future Large Synoptic
Survey Telescope\footnote{\tt www.lsst.org} or the imaging
component of the ESA/Euclid\footnote{\tt www.euclid-imaging.net} satellite.

Redshift space distortions (RSD) \cite{Kai,hamilton98}
can be used to  understand the (linear) growth of
structures, which provides a direct path to study the origin 
of cosmic acceleration. On large scales, RSD
arises from the coherent velocities of galaxies and reveals how
perturbations grow in time. Typically this
method requires measuring of galaxy clustering in 3 dimensions (3D) in order to sample
directions parallel and transverse to
the line-of-sight where the effect is maximized or cancels out
completely (see
e.g. \pcite{okumura08,guzzo09,cabre09,blake11,reid12,kazin13}
and references therein).

Over the past few years it has been however shown that the effect of RSD 
 is also present, albeit with a smaller contribution, in the
angular (2D) clustering of photometric galaxy samples if they are selected in 
  photometric redshift bins (see for instance
\pcite{2010MNRAS.407..520N,2011MNRAS.414..329C,2011MNRAS.415.2193R}). This concrete
idea has been already applied to data using a sample of photometric
Luminous Red Galaxy (LRG, see \pcite{blake07,nikhil,2011arXiv1104.5236C,thomas2011}).

Yet all the previous studies focused on the angular clustering 
from a set of measurements of auto-correlation in one or more redshift
bins.  In turn cross-correlations have been proposed and
mostly used to test for different systematics and to calibrate redshift
distributions (see for instance \pcite{newman08,thomas2011,benjamin2013}).

Hence the goal of this paper is, on the one hand, to extend these analysis to include also
 the cross-correlations between redshift bins in order to account for
 some radial information. This is motivated by the recent findings of
\pcite{PhysRevD84063505,1206.3545,Asorey2012} who show how a tomographic (2D) study 
 involving auto and cross correlations can yield similar constrains on
 cosmological parameters as a full spatial (3D) study. It is also important because
 a 2D formalism can naturally combine redshift space
 distortions with weak lensing \cite{1112.4478,1109.4852,dePutter2013,kirk2013}. This is particularly relevant
 to discriminate between different models of modify gravity
 and general relativity by breaking the degeneracies between expansion history
 and growth of structure. 

 On the other hand we will also investigate
 the improvements brought by considering two different populations (and their cross
 correlations) in the likelihood analysis for the growth rate.
 This is motivated by the fact that for the
 spectroscopic analysis, the combination of different samples tracing
 the same underlying matter fluctuations can be used to decrease
 sampling variance and improve considerably the constrains
 in growth of structure \cite{McDonaldSeljak,white2009,gil-marin2010}.	

This paper is organized as follows. In Sec.~\ref{sec:methodology} we
lay out the methodology, including the analytical tools, the definition
of the different samples and surveys and the likelihood analysis. In Sec.~
\ref{sec:results} we present our result, and in
Sec.~\ref{sec:conclusions} our conclusions.

\section{Methodology}\label{sec:methodology}

Our goal is to study the effect of RSD in angular clustering,
especially its usefulness to derive constrains on the growth of structure at
large scales. We study angular clustering using auto- and cross-
correlations between redshift bins. The inclusion of cross
correlations between different radial shells allow us to include the
radial modes that account for scales comparable to the bin
separation. On the other hand, the  angular spectra of each redshift shell includes information mainly from transverse modes. 

With the idea of a potential sample variance mitigation in the analysis, we also
consider the correlation between the angular clustering of
different tracers of matter, considering them either independent
(i.e. each tracer in a
different patch of the sky) or correlated (same sky).

Throughout this paper we use 
\camb\footnote{camb.info/sources} \cite{cambt,lewis2007,2011arXiv1105.5292C},
including cross correlations between radial bins and the correlations
between different populations. Let us note that we use the exact
$C_{\ell}$ computation in \camb, because in angular clustering
the imprint of redshift distortions affect mainly the largest scales, which are not 
included when using the Limber approximation
\cite{1954ApJ...119..655L,LoVerde2008,2011MNRAS.414..329C}. Moreover
the Limber approximations does not account for clustering in adjacent
redshift bins.

\subsection{Fiducial survey and galaxy samples}
\label{sec:survey}

We start by describing the fiducial photometric survey that we assume in
our analysis (characterized by a redshift range and a survey
area) and the different galaxy samples considered within that volume
(characterized by the bias $b$, the accuracy of photometric redshift
estimates $\sigma_z$ and their redshift distribution). 

Our fiducial survey is similar to the full DES, with an
area coverage of one octant of the sky (i.e., $f_{sky}=1/8$) and a
redshift range $0.4<z<1.4$. We characterize  the redshift distribution of galaxies within this
survey by
\begin{equation}
\frac{dN_\alpha}{dzd\Omega}=N_{gal}^{\alpha}  \left(\frac{z}{0.5}\right)^{2}e^{-\left(\frac{z}{0.5}\right)^{1.5}}
\label{eq:galaxyselectionfunction}
\end{equation}
where $N_{gal}^{\alpha}$ is a
normalization related to the total number of galaxies of each
population sample, denoted by $\alpha$. We typically consider two
types of sample populations, one with bias $b=1$ and 
$\sigma_z=0.05(1+z)$ (Pop1) and another with $b=2$ and
$\sigma_z=0.03(1+z)$ (Pop2) \cite{Banerji2008,2011MNRAS.415.2193R}. For simplicity we
consider the same redshift distribution for all samples with a
fiducial comoving number density of
$n(z=0.9)= 0.023 {\it
  h}^3{\rm Mpc}^{-3}$, unless otherwise stated.  This value corresponds to a total of $\sim
3\times 10^8$ galaxies within the survey redshift range and matches
the nominal number galaxies expected to be targeted above the magnitude
 limit of DES ($i<24$). For more details about DES specifications we
 refer the reader to \pcite{deswhitepaper}. 


In Table \ref{table:BinConfigurationDES} we show the different redshift binning schemes in which we divide our survey prior to study the clustering either with the auto-correlations or with the 2D tomography that also includes the cross-correlations between bins.
\begin{table}
{\center
\begin{tabular}{cc}
Number of bins $N_z$ & $\Delta z/(1+z)$ \\ 
\hline
4&0.15  \\
6&0.1  \\
8&0.08 \\
12&0.05  \\
19&0.03 \\
\hline  \\
\end{tabular}
\caption{The different redshift bin configurations considered in our paper, within
  a photometric redshift range of $0.4<z<1.4$. We show the total number of
  bins and their redshift width $\Delta z$ (which evolves with
  redshift in the same manner as the photo-z).}
\label{table:BinConfigurationDES}
}
\end{table}
Note that we consider consecutive bins with an evolving bin width with redshift, 
  i.e. $\Delta z \propto (1+z)$, to match the photometric uncertainty which also assumes a linear evolution
 with redshift.

\subsection{Angular power spectrum}
\label{sec:2Dpower}

In our analysis we study angular clustering using the angular power spectrum of the projected overdensities in the space
of spherical harmonics. The auto-correlation power spectrum at redshift bin $i$, for a single population, is given by:
\beq
C_\ell^{i i} = \frac{2}{\pi} \int dk\; k^2 P_0(k) \left(\Psi_l^{i}(k)+\Psi^{i,r}_l(k)\right)^2
\label{eq:cl}
\eeq
where
\beq
\Psi_\ell^i(k) = \int dz \; \phi_i(z) b(z) D(z) j_\ell(k r(z))
\label{eq:psi}
\eeq
is the kernel function in real space and
\bea
\Psi^{i,r}_\ell(k)&=&  \int dz \; \phi_i(z) f(z) D(z) \left[ \frac{2l^2+2l-1}{(2\ell+3)(2\ell-1)} j_\ell(kr)
\right. \nonumber \\
&-& \left.\frac{\ell(\ell-1)}{(2\ell-1)(2\ell+1)} j_{\ell-2}(kr)\right.\nonumber\\
&-&\left. \frac{(\ell+1)(\ell+2)}{2\ell+1)(2\ell+3)} j_{\ell+2}(kr) \right].
\label{eq:RSDpsi}
\eea
should be added to $\Psi^{i}_{\ell}$ if we also include the linear
Kaiser effect \cite{fisher94,nikhil}. In Eqs.~(\ref{eq:psi},\ref{eq:RSDpsi}) $b(z)$ is the bias
(assumed linear and deterministic), $D(z)$ is the linear growth factor and
$f(z)\equiv\partial \ln D / \partial \ln a$ is the growth rate.
Photo-z effects are included through the radial selection function $\phi(z)$, see below. 

For the case of 1 population, there are $N_z$ auto-correlation spectra, one per radial bin. Then, we add to  our observables the $N_z(N_z-1)/2$ cross-correlations between different redshift bins. These are given by
\beq
C_\ell^{ij} = \frac{2}{\pi} \int dk\; k^2 P(k) \left(\Psi_\ell^i(k)+\Psi^{i,r}_\ell(k)\right) \left(\Psi_\ell^j(k)+\Psi^{j,r}_\ell(k)\right)
\label{eq:clcross}
\eeq
Therefore, we are considering $N_z(N_z+1)/2$ observable angular power spectra when reconstructing clustering information from tomography using $N_z$ bins, for a single tracer.

If we combine the analysis of two tracers, $\alpha$ and $\beta$, the angular power spectrum is given by
\bea
C_\ell^{i_\alpha j_\beta} &=& \frac{2}{\pi} \int dk\; k^2 P(k) \left(\Psi_\ell^{i_{\alpha}}(k)\right.\nonumber \\
&+&\left.\Psi^{i_\alpha,r}_\ell(k)\right)  \left(\Psi_\ell^{j_\beta}(k)+\Psi^{j_\beta,r}_\ell(k)\right),
\label{eq:clcross2pop}
\eea
where $\Psi_\ell^{i}$ and $\Psi_\ell^{i,r}$ characterize each
galaxy sample through the radial selection function $\phi_i(z)$ and the bias $b(z)$ in
expressions (\ref{eq:psi}) and (\ref{eq:RSDpsi}) . 
We use the general notation where $C_\ell^{i_\alpha j_\beta}$ is the correlation between redshift bin $i$ of population $\alpha$ with redshift bin $j$ of population $\beta$. By definition,
\bea
C_\ell^{i_\alpha j_\beta} &=& C_\ell^{j_\beta i_\alpha } \\
C_\ell^{i_\alpha j_\beta} &\neq& C_\ell^{j_\alpha i_\beta }\ \ \hbox{for}\ \ \alpha \neq \beta; i\neq j
\eea
Then the total number of observables is $2N_z(2N_z+1)/2$ if we consider the same redshift bins configuration for both populations, in the case in which both are correlated.

\begin{figure*}
\begin{center}
\includegraphics[width=0.5\textwidth]{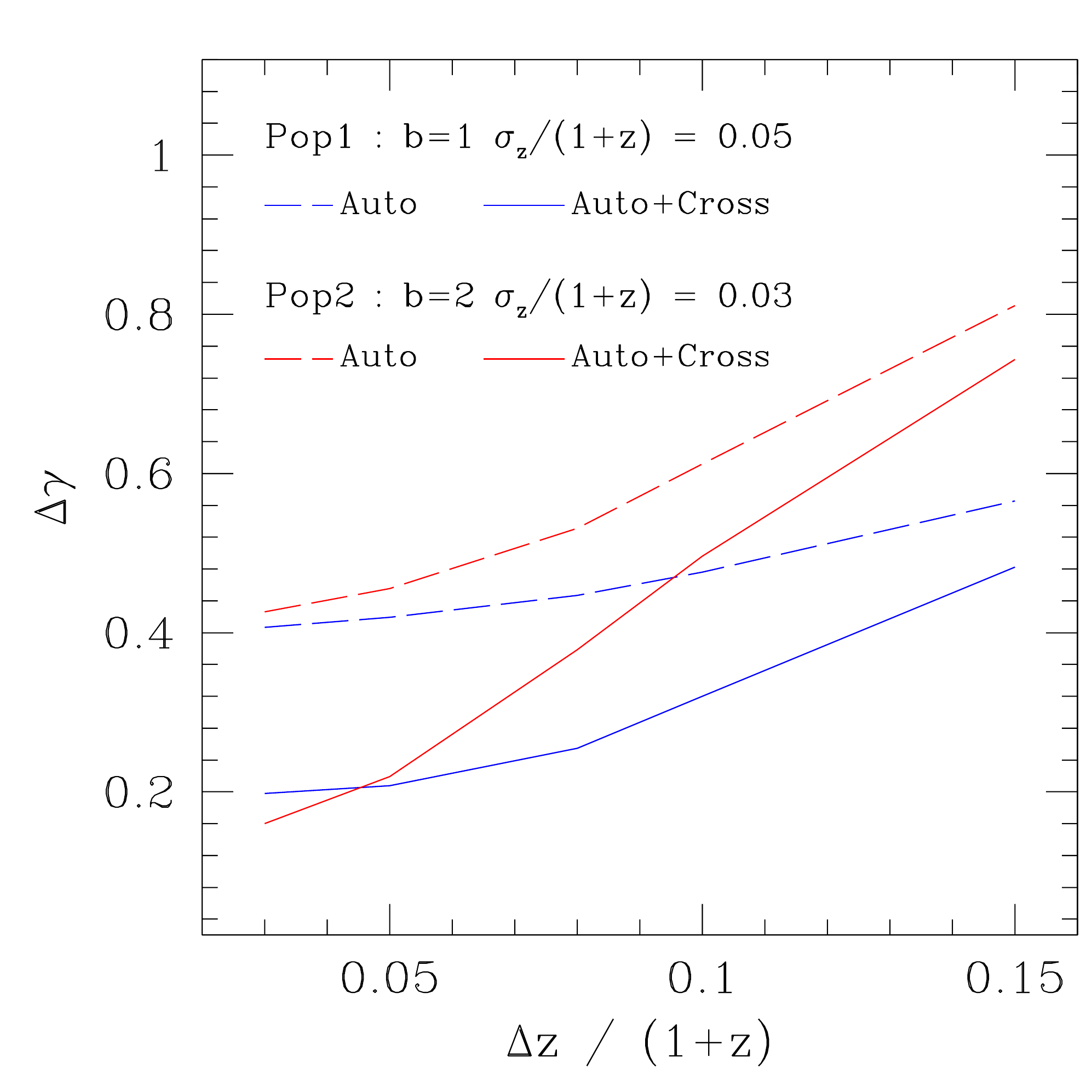}
\caption{{\it The gain from adding redshift-bins
    cross-correlations}. Dashed lines show the expected 1-$\sigma$
  constrains in $\gamma$ from the
  combined analysis of angular auto-correlation in photo-z bins spanning $0.4 < z <
  1.4$, as a function of the  bin width $\Delta z/(1+z)$ (see
  Table~\ref{table:BinConfigurationDES} for the corresponding total number of bins). Different
  colors correspond to different populations with bias and $\sigma_z$
as labeled. Solid lines show,
  for each population, the same study but also including all the 
  cross-correlations between bins (and their complete covariance). For optimal bin widths $\Delta z
  \lesssim \sigma_z$ the gain from including 
  cross-correlations is $\sim 2$ or better.}
\label{fig:fig1}
\end{center}
\end{figure*}

\subsubsection{Radial selection functions}

The radial selection functions $\phi_i$ in Eqs.~(\ref{eq:cl},\ref{eq:clcross},\ref{eq:clcross2pop}) encode the probability to include a galaxy in the given redshift bin. Therefore, they are the product of the corresponding galaxy redshift distribution and a window function that depends on selection characteristics (e.g binning strategy),
\beq
\phi_i^\alpha(z)=\frac{dN_\alpha}{dz}\, W_i(z)
\eeq
where $dN_\alpha/dz$ is given by
Eq.~(\ref{eq:galaxyselectionfunction}). We include the fact that we
are working with photo-z by using the following window function:
\beq
W_i(z)=\int{dz_pP(z|z_p)W^{ph}_i(z_p)},
\eeq
where $z_p$ is the photometric redshift and $P(z|z_p)$ is the
probability of the true redshift to be $z$ if the photometric estimate
is $z_p$. For our work we assume a top-hat selection $W^{ph}_i(z_p)$ in photometric redshift and that $P(z|z_p)$ is Gaussian with standard deviation $\sigma_z$. This leads to,
\beq
\phi^\alpha_i(z)\propto \frac{dN_\alpha}{dz} \left( {\rm erf}\left[\frac{z_{p,max}-z}{\sqrt{2}\sigma^\alpha_z}\right]-{\rm erf}\left[\frac{z_{p,min}-z}{\sqrt{2}\sigma^\alpha_z}\right]\right)
\label{eq:nz}
\eeq
where $z_{p, min}$ and $z_{p, max}$ are the (photometric) limits of each redshift bin considered and
$\sigma^\alpha_z$ is the photometric redshift error of the given
population $\alpha$ at the corresponding redshift.

\subsubsection{Covariance matrix of angular power spectra}
We assume that the overdensity field is given by a Gaussian distribution and therefore, the covariance between correlation  $C_\ell^{i_\alpha j_\beta}$ and correlation $C_\ell^{p_\alpha q_\beta}$ is given by
\beq
{\rm Cov}_{\ell,(i_\alpha j_\beta)(p_\mu q_\nu)}=\frac{C_{\ell}^{obs, i_\alpha p_\mu} C_{\ell}^{obs, j_\beta q_\nu}+
  C_{\ell}^{obs, i_\alpha q_\nu} C_{\ell}^{obs, j_\beta p_\mu}}{N(l)}
\label{eq:CovObs2D}
\eeq
where $N(\ell)=(2\ell+1)\Delta\ell f_{sky}$ is the number of
transverse modes at a given $\ell$ provided with a bin width $\Delta
\ell$. We set $\Delta \ell = 2/f_{sky}$, the typically chosen value 
to make Cov
block-diagonal \cite{cabre07,2011MNRAS.414..329C}. 
In this case, bins in $\ell$ are not correlated between them.

Therefore, for each $\ell$ bin, we define a matrix with $2N_z(2N_z+1)/2$ rows, where $N_z$ is the number of redshift bins, taking into account  the covariances and cross-covariances of auto and cross-correlations between each population and among them.
In order to include observational noise we add to the auto-correlations of each population in Eq.~(\ref{eq:CovObs2D}) a shot noise term
\beq
C_{\ell}^{obs, i_\alpha j_\beta}=C_{\ell}^{i_\alpha j_\beta} +\delta_{i_\alpha j_\beta}\frac{1}{\frac{N_{gal}(j_\beta)}{\Delta\Omega}}
\eeq
that depends on the number of galaxies per unit solid angle included in each radial bin.
We define the $\chi$ assuming the observed power spectrum $C_{\ell}^{obs}$ corresponds to our fiducial cosmological model discussed in Sec.~(\ref{sec:cosmologicalmodel}), while we call $C^{mod}_{\ell}$ the one corresponding to the cosmology being tested,
\beq
\chi^2=\sum_{\ell} \left(C_{\ell}^{obs}-C_{\ell}^{mod}\right)^{\dagger}
{\rm Cov}_{\ell}^{-1}\left(C_{\ell}^{obs}-C_{\ell}^{mod}\right).
\label{eq:2Dchi2}
\eeq

\subsection{Cosmological model and growth history}
\label{sec:cosmologicalmodel}

Throughout the analyses, we assume the underlying cosmological model to be a flat $\Lambda$CDM with cosmological parameters $w=-1$, $h=0.7$, $n_s=0.95$, $\Omega_{m}=0.25$, $\Omega_b=0.045$ and $\sigma_8=0.8$.
These parameters specify the cosmic history as well as the linear spectrum of fluctuations $P_0$.
In turn, the growth rate can be well approximated by,
\beq
f(z) \equiv \Omega_m(z)^\gamma
\label{eq:fz}
\eeq
and $\gamma = 0.545 $ for $\Lambda$CDM. Consistently with this we obtain the growth history as
\beq
D(z) \equiv \exp \left[ - \int_0^z \frac{f(z)}{1+z} dz    \right]
\eeq
(where $D$ is normalized to unity today). The parameter $\gamma$ is
usually employed as an effective way of characterizing modified
gravity models that share the same cosmic history as GR but different
growth history \cite{linder1}. Our fiducial model
assumes the GR value $\gamma = 0.545$.
In order to forecast the constrains on $\gamma$
we consider it as a free parameter independent of redshift.

With these ingredients, we do a mock likelihood sampling in which we
assume that the theoretical values for the correlations at the
fiducial value of the parameters corresponds to the best fit
position. The likelihood is based on the $\chi^2$ given in
(\ref{eq:2Dchi2}). In our case, we keep fixed all the parameters and
only allow $\gamma$ to vary, and then we estimate 68\% confidence
limits of it. In the case in which we show constrains on $f\sigma_8$,
we vary this quantity (that now depends on redshift, thus the number
of fitting parameters is a function of the bin configuration), fixing the rest of parameters. The maximum
$\ell$ considered in the analysis is
$\ell_{max}=r(\bar{z}_{Survey})k_{max}\sim 220$ for $k_{max}=0.1\, {\it
  h} {\rm Mpc}^{-1}$, while for the largest scales we set
$l_{min}=2$. We had to adapt \camb \ in order to constrain $\gamma$ or
$f\sigma_8$ using the same technique described in the Appendix A of \pcite{Asorey2012}.

\begin{figure}
\begin{center}
\includegraphics[trim = 0cm 2.5cm 0cm 5cm, clip=true, width=0.43\textwidth]{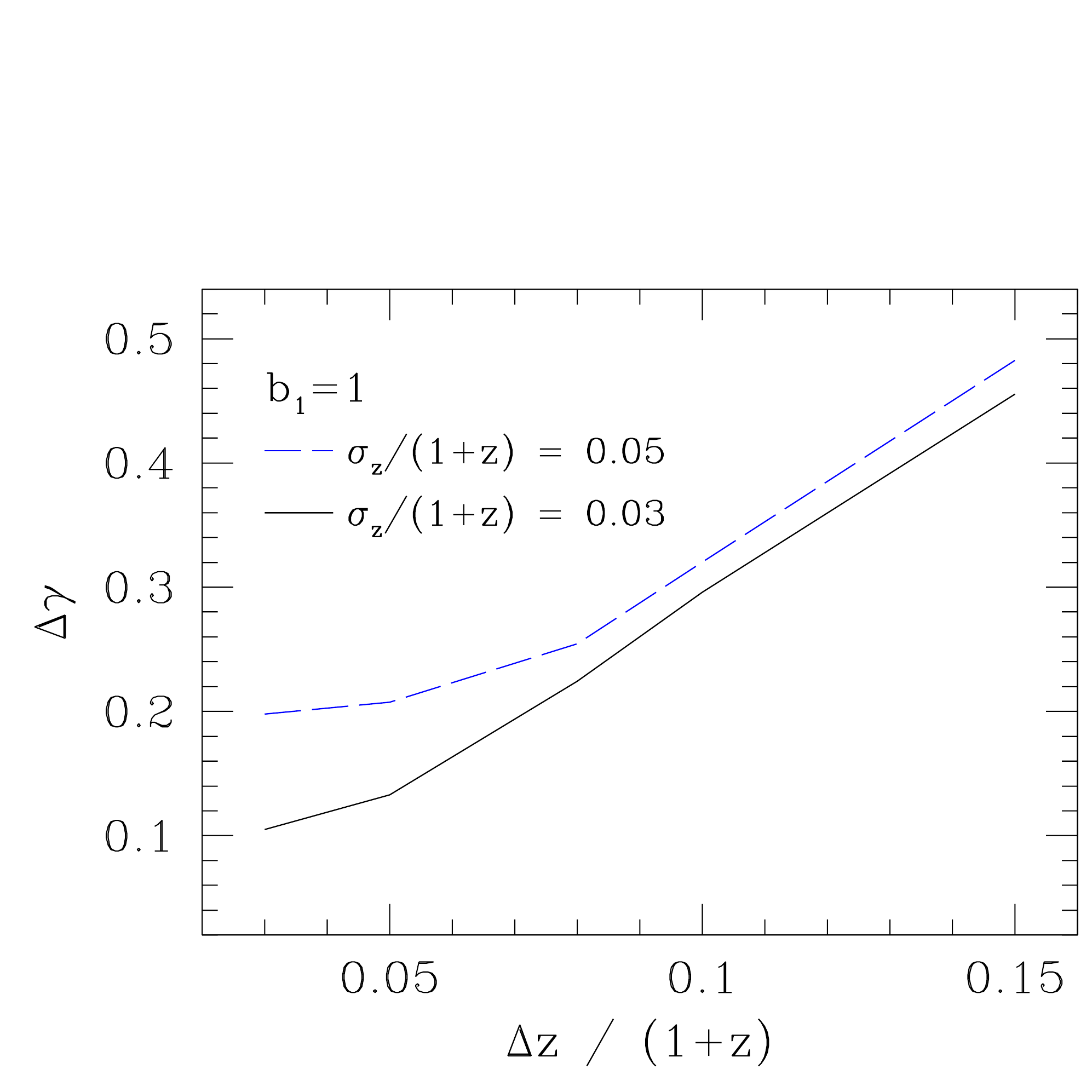} \\
\includegraphics[trim = 0cm 0cm 0cm 5cm, clip=true, width=0.43\textwidth]{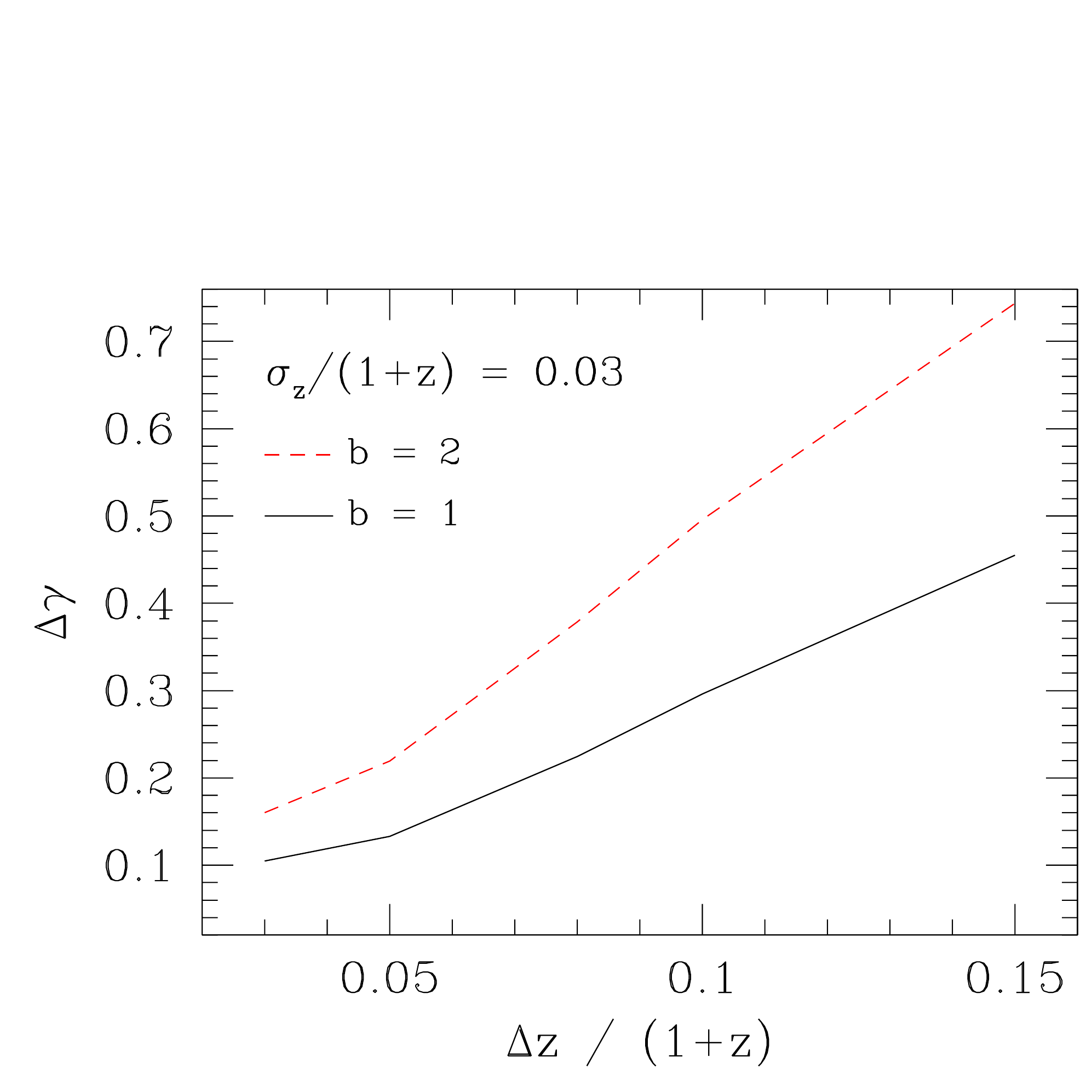}
\caption{
Dependence on photo-z (top panel) and bias (bottom panel) for a
    one-population constrains in $\gamma$, as a function of bin
    width (same as in Fig.~\ref{fig:fig1}). The panels show that lower $b$ and/or lower $\sigma_z$
    yields better constrains in $\gamma$. This is hence 
   a competing interplay because lower $b$ would correspond to a
   fainter sample with typically worse photometric errors. The
   dependence on bias seems however slightly stronger.}
\label{fig:fig2}
\end{center}
\end{figure}

\section{Results}
\label{sec:results}

In this section we discuss the constrains on the growth index, $\gamma$
defined in Eq.~(\ref{eq:fz}) as obtained for the different redshift
bin configurations of Table \ref{table:BinConfigurationDES}. First of
all, we study how well we can determine $\gamma$ using different
single galaxy populations but including as observables also the cross
correlation between bins (for a given single population). We also study how
the constrains depend on the bias and in the photometric redshift accuracy
of the different galaxy samples.
Then, we study the precision achievable when one combines different
tracers in the analysis and how this depends on bias, photo-z and in
particular, the shot-noise level of the sample.

Lastly we discuss the constrains that we obtain when looking into the more
standard $f(z)\sigma_8(z)$ as a function of redshift, and consider
auto and cross-correlations of one or two galaxy samples.

\subsection{Redshift-space distortions with a single photometric population}
\label{sec:singlepop}

Let us first consider the constrains on the growth index using single
photometric populations. Figure \ref{fig:fig1} shows the
1-$\sigma$ errors expected on $\gamma$ from a combined analysis of all the
consecutive photometric redshift bins in the redshift range $0.4 < z <
1.4$ as a function of the bin width (i.e. each of the configurations detailed in Table 1)\footnote{Note that different redshift bins can be strongly
  correlated depending on bin width and photo-z. We do include this
  covariance.}.

In red we show the constrains on $\gamma$ corresponding to an LRG-type
sample, with bias $b=2$ and a photometric redshift
$\sigma_z/(1+z)=0.03$ (Pop2). 
Blue lines correspond to an unbiased population with
$\sigma_z/(1+z)=0.05$ (Pop1).

Dashed lines correspond to the case in which we only use
the auto-correlations in each redshift bin while solid lines
corresponds to the full 2D analysis that includes all the
cross-correlations in our vector of observables. Recall than in the
first case the cross-correlations are included in the covariance
matrix of the auto-correlations (but not as observables). 
We see that constrains from a full 2D analysis, including auto and
cross-correlations are a factor $\sim 2$ or more better than those
from using only auto-correlations.

From Fig. ~\ref{fig:fig1}, it is clear that in all cases the bin configuration can be
optimized, with the best results obtained when $\Delta z \sim
\sigma_z$. In addition, there is a competing effect between $\sigma_z$
and bias. For broad bins ($\Delta z \gg \sigma_z$) the photo-z of the
populations is masked in the projection and the bias dominates the
$\gamma$ constrains. Smaller bias gives more relevance to RSD and
better $\gamma$ constrains. As one decreases the bin width the
population with better photo-z  (typically the brighter, with higher bias), denoted Pop2,
allows a more detailed account of radial modes improving the derived
errors on $\gamma$ more rapidly than Pop1 until they become slightly
better. This optimization is possible until one eventually reaches bin
sizes comparable to the corresponding photo-z (what sets an
``effective'' width) and the constrains flatten out.

In Fig.~\ref{fig:fig2}, we study in more detail the dependence of
constrains with respect to galaxy bias $b$ and photo-z accuracy. In
the top panel of Fig.~\ref{fig:fig2} we show standard deviation of the
growth index, $\Delta \gamma$, fixing the sample bias to $b=1$ and allowing two
values for photo-z accuracy. Red line represents a sample in which
$\sigma_z/(1+z)=0.05$ while blue line has an error of
$\sigma_z/(1+z)=0.03$. In both cases the constrain flattens once $\Delta z
\sim \sigma_z$ and the optimal error improves roughly linearly with $\sigma_z$.
The dependence on the linear galaxy bias, $b$, is shown in the bottom
panel of Fig.~\ref{fig:fig2} (for fixed $\sigma_z$). We see that the constrains degrade
almost linearly with increasing bias (see also \pcite{2011MNRAS.415.2193R}).  As
  discussed before, this is because the
lower the bias the larger the relative impact of RSD, which results in 
better constraints on $\gamma$.
\begin{figure*}
\begin{center}
\includegraphics[width=0.5\textwidth]{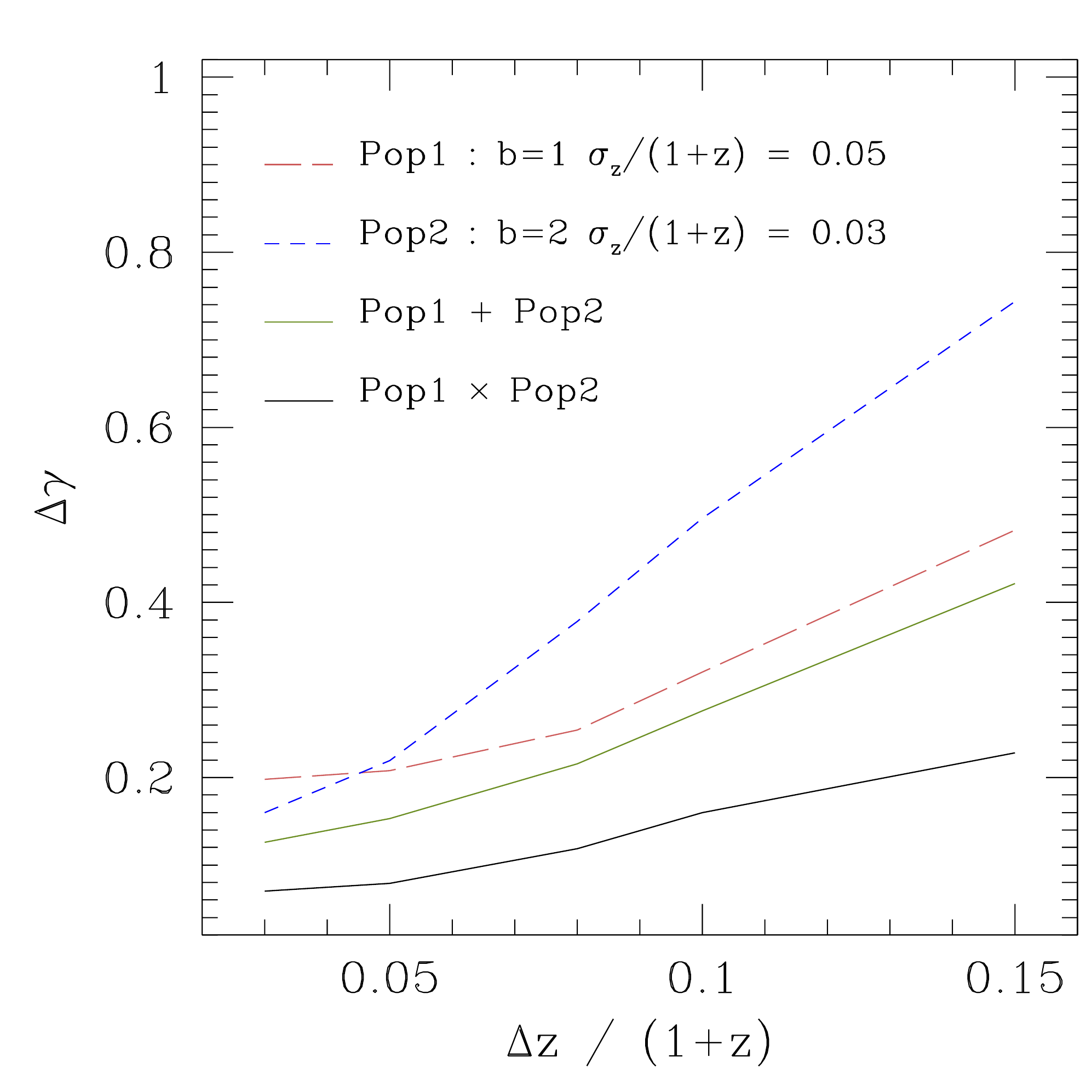}
\caption{
{\emph{The gain from combining galaxy populations}: Comparison
  of the 68\% standard deviations in the growth index from single
  population analysis (dashed lines) with respect to the combined
  analysis of these two
  populations over the same field (black solid), using all the angular
  auto and
  cross-correlations.  Remarkably the combination yields errors at least 2 times better than
  any of the single population cases. 
  The solid green line corresponds to the combination of the two
  samples assuming they are independent (i.e. from
  different parts of the sky). 
  As shown, the combination of 
  correlated populations (same sky) yield stronger constrains than any other case.}}
\label{fig:fig3}
\end{center}
\end{figure*}

In summary we have shown that using the whole 2D tomography
(auto+cross correlations) allows considerable more
precise measurements of $\gamma$, a factor of 2 or better once the
bin width is optimal for the given sample. Hence in what follows we concentrate in full
tomographic analysis.

\subsection{Redshift-space distortions with 2 photometric populations}
\label{sec:twopops}

We now turn to an analysis combining two galaxy populations
as two different tracers of matter. In Figure \ref{fig:fig3} we
compare the constrains from single tracers with respect to the
combination of both. As before the populations used in the comparison
correspond to $b=1$ and $\sigma_z/(1+z)=0.05$ (Pop
1) and a population with $b=2$ and $\sigma_z/(1+z)=0.03$ (Pop
2). Their respective constrains in $\gamma$ are the dashed red and
blue lines (same as solid lines in Fig.~\ref{fig:fig1}).

If we combine both tracers and their cross-correlation in the same analysis we obtain the constrains
given by black solid line, notably a factor of $2-3$ better
compared to the optimal single population configuration. 

In order to understand how much of this gain is due to ``sample variance
cancellation'', in analogy to the idea put forward in \pcite{McDonaldSeljak},
we also considered combining the two samples assuming they are located in
different parts of the sky (and hence un-correlated). We call this case Pop1+Pop2 in
Fig.~\ref{fig:fig3} (solid green line). In such
scenario the total volume sampled is the sum of the volumes sampled by
each population (in our case, two times the full volume of
DES). This explains the gain with respect to the single population
analysis. Nonetheless, the ``same sky'' case Pop1 $\times$ Pop2 (where cosmic
variance is sampled twice) still yields better constrains, a factor of
$\sim 1.5 - 2$, even though the area has not increased w.r.t. Pop1 or Pop2 alone.

\begin{figure}
\begin{center}
\includegraphics[trim = 0cm 0cm 0cm 4cm, width=0.45\textwidth]{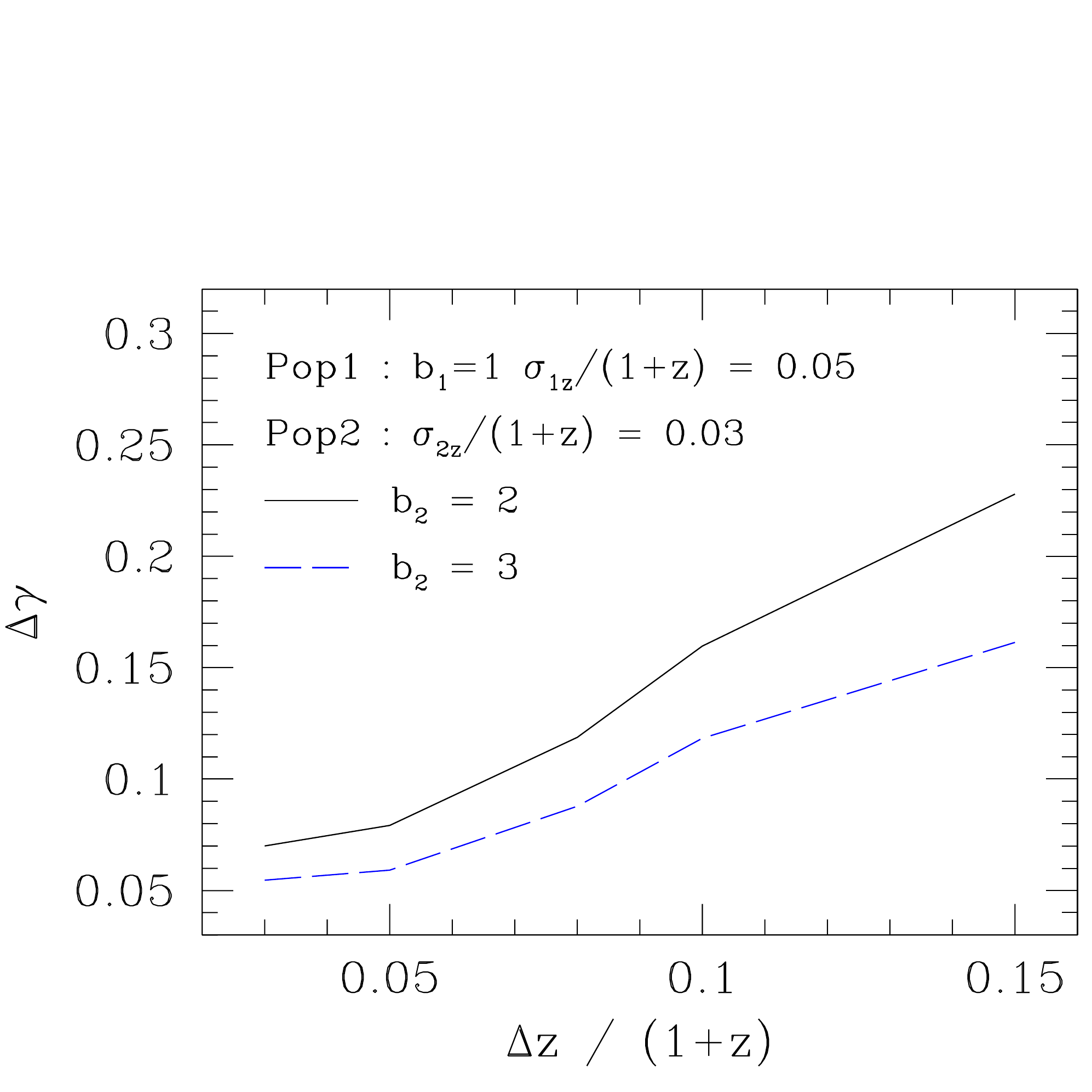}
\caption{\emph{Dependence on bias}. Increasing the bias difference
  between the samples improves the constrains on $\gamma$. The solid black
  line corresponds to the combination Pop1 $\times$ Pop2 of a highly biased sample such
  as LRGs (Pop2) with an unbiased one (Pop1), while the blue dashed to
  cluster-like bias tracer as Population 2.}
\label{fig:fig4a}
\end{center}
\end{figure}

\begin{figure}
\begin{center}
\includegraphics[trim = 0cm 0cm 0cm 4cm, width=0.45\textwidth]{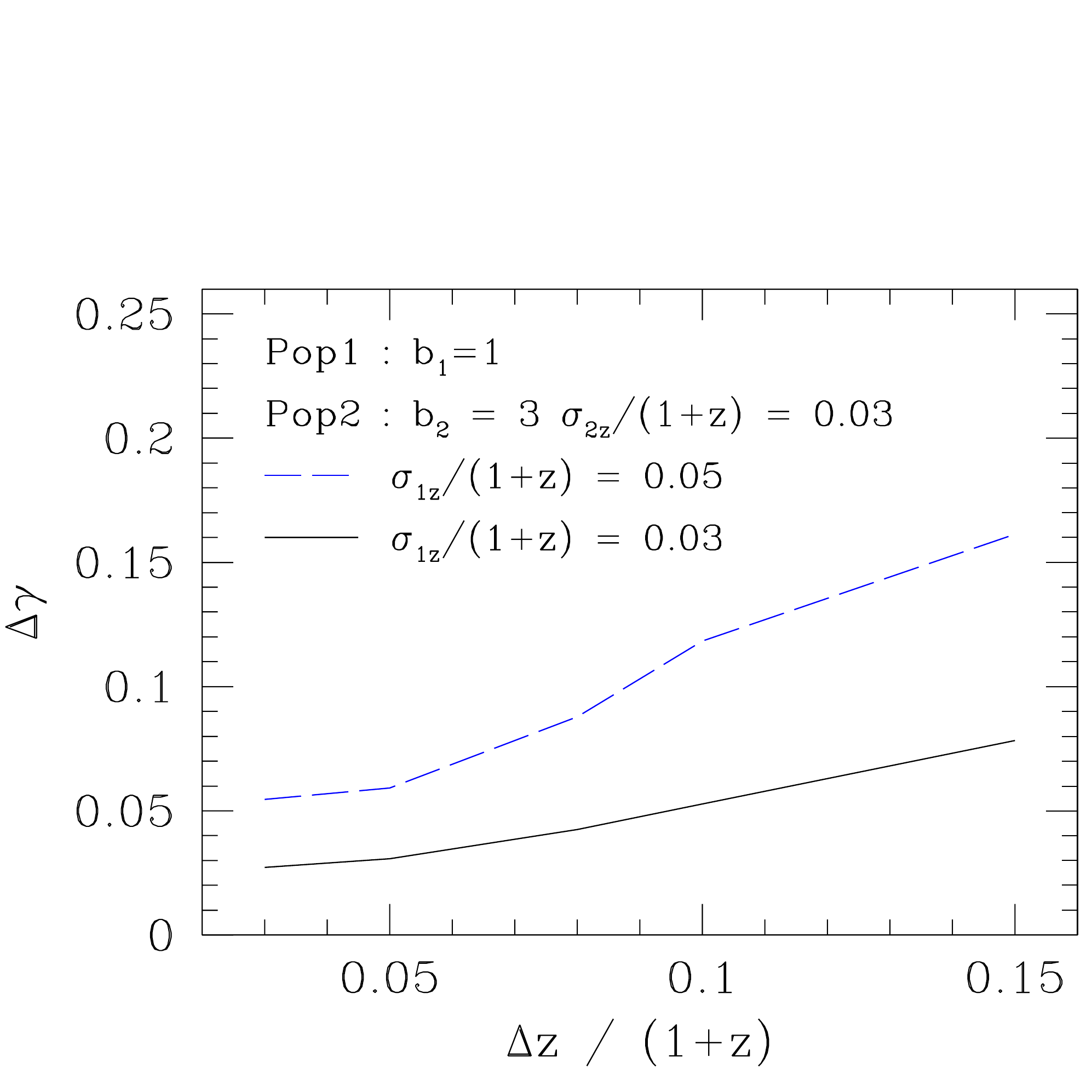}
\caption{ 
\emph{Dependence on photometric redshift error}. Similar to Fig.~\ref{fig:fig4a}  but now changing the photo-z of the unbiased
  sample (Pop1) for a fixed 2nd population. The error on $\gamma$ depend roughly linear with
  $\sigma_z/(1+z)$ for optimal bin widths.}
\label{fig:fig4b}
\end{center}
\end{figure}

In all, the total gain of a full 2D study with two populations
(including all auto and cross correlations in the range $0.4<z<1.4$) 
w.r.t. the more standard analysis with a single population
and only the auto-correlations in redshift bins (dashed lines of Fig.~\ref{fig:fig1}) can reach a factor of $\sim 5$.
\begin{figure}
\begin{center}
\includegraphics[trim = 0cm 0cm 0cm 4cm, width=0.43\textwidth]{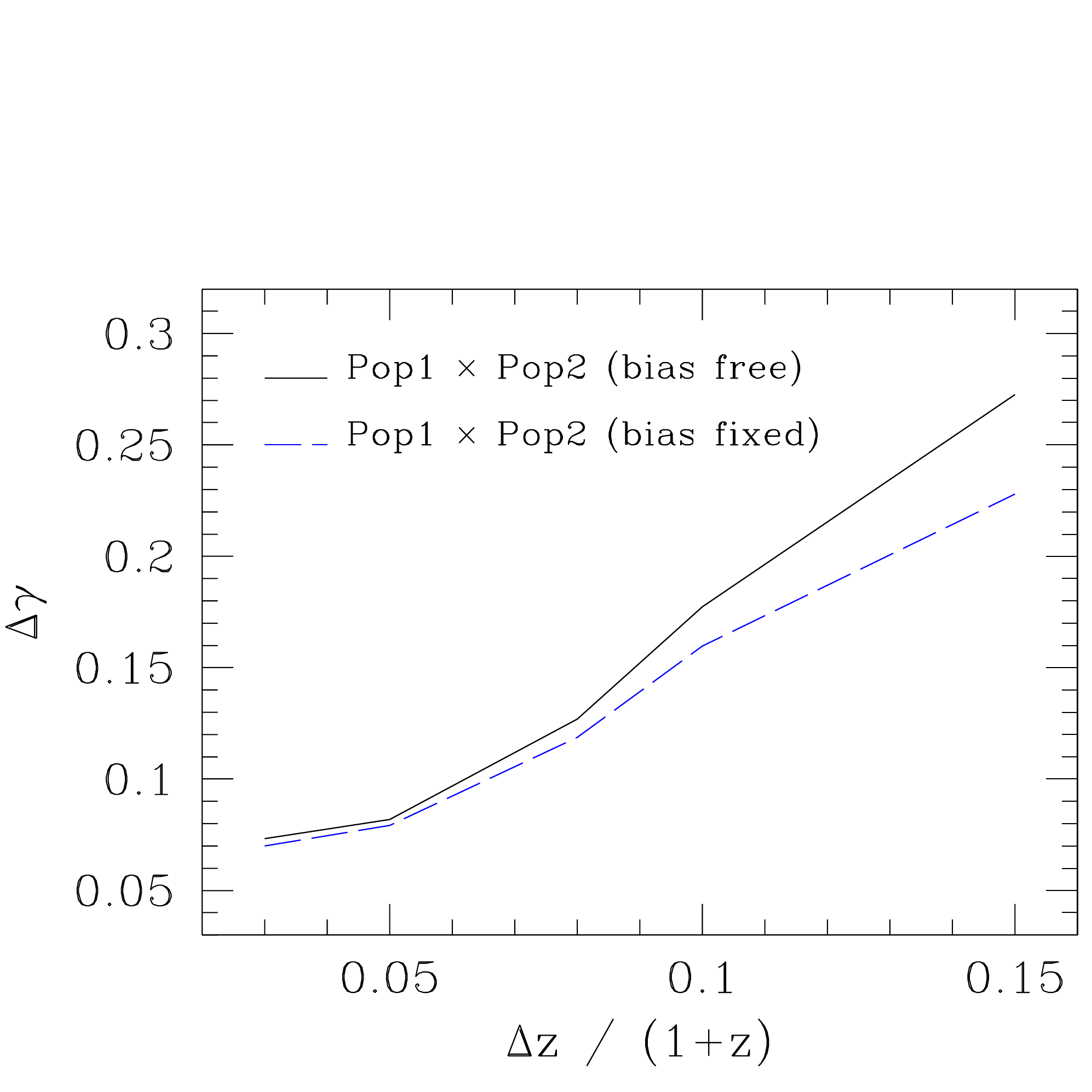}
\includegraphics[trim = 0cm 0cm 0cm 4cm, width=0.43\textwidth]{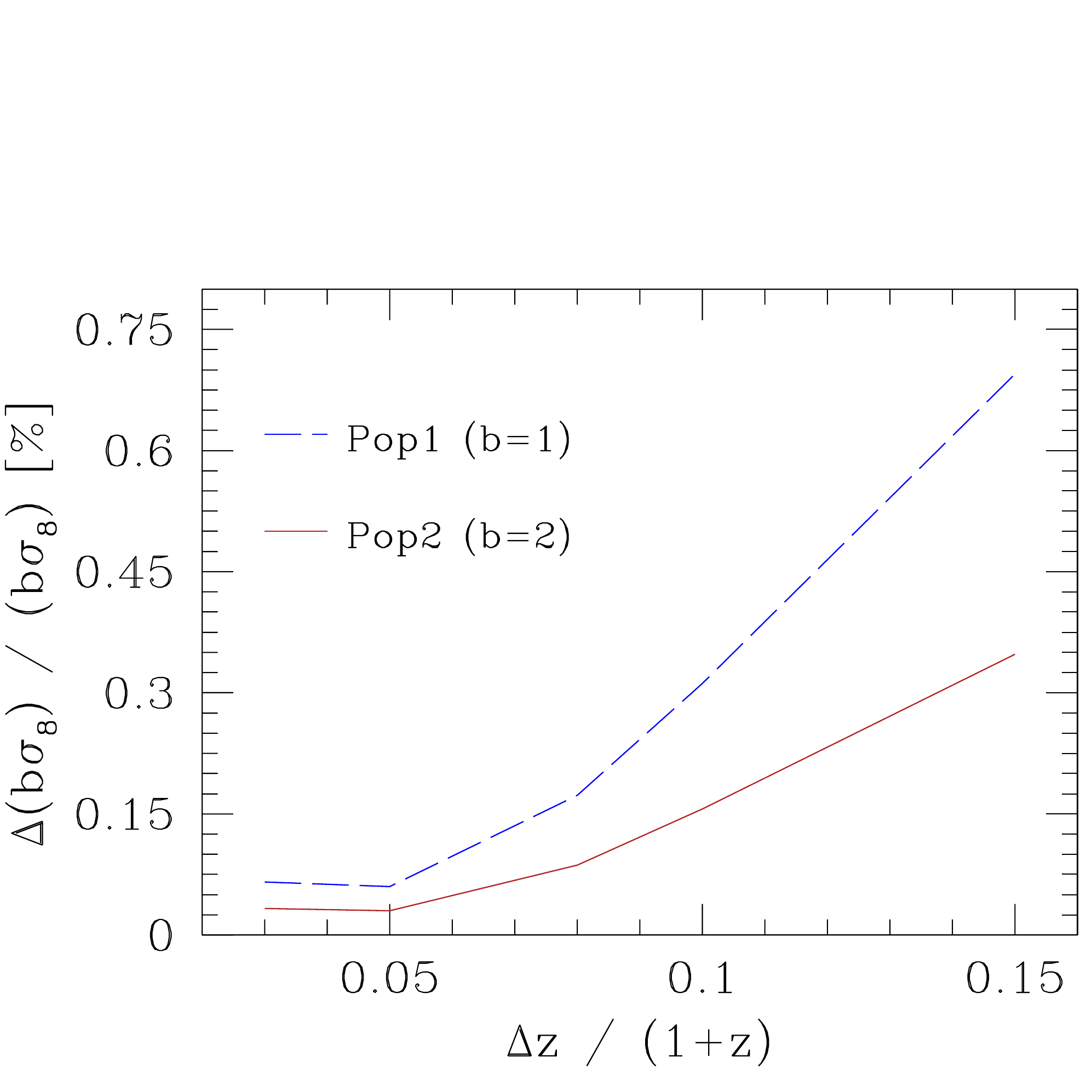}
\caption{
\emph{Bias free case}: If the biases of the samples are free
parameters to marginalize over we find that constrains on $\gamma$ degrade only slightly compared with the
bias fixed case. In particular for the thinner
redshift bins configurations. This is because biases are determined with
relative errors smaller than 1\% (bottom panel).}
\label{fig:fig3b}
\end{center}
\end{figure}

As a next step we show how the combined analysis of two tracers
depends on the relative difference on the bias and photo-z errors of
the populations. In Fig.~\ref{fig:fig4a} we keep Pop1 fix
(with $b=1$ and $\sigma_z/(1+z)=0.05$) and we vary the bias of Pop2
from  $b=2$ (LRG type bias) to $b=3$ (galaxy clustering like). We
keep $\sigma_z/(1+z)=0.03$ fixed for Pop2.
As expected, increasing the bias 
difference between the samples
improves the constrains on $\gamma$ in a roughly linear way.

If we now have an unbiased tracer and a highly biased one with $b=3$,
while both tracers have the same $\sigma_z/(1+z)=0.03$ we obtain
constrains given by the black line in Fig.~\ref{fig:fig4b}. Those
constrains are better than the case in which the unbiased galaxies
photo-z is worse, $\sigma_z/(1+z)=0.05$ (given by the dashed blue line). Therefore, if we
determine photometric redshifts of the unbiased galaxies with higher
accuracy we will be able to measure the growth rate with higher precision.

One caveat so far is that we have always assumed that biases are
perfectly known (bias fixed). Hence, in the top panel of Fig.~\ref{fig:fig3b} we show
how the constrains on $\gamma$ change if we instead consider them as
free parameters and marginalize over. We see that the difference is very small, in particular
once the bin configuration is optimal. The reason for this is clear from
the bottom panel that shows the relative error obtained for the bias
of each sample in the bias free case. Because the bias is so well
determined (sub-percent) the marginalization over them does not impact
the error on $\gamma$.

\subsubsection{The impact of photo-z uncertainties}
 
The constrains on growth of structure presented in this paper rely to
a good extent on cross-correlations between redshift bins, in turn largely
determined by the overlap of the corresponding redshift distributions.
So far we have assume a perfect knowledge of these distributions,
given by Eq.~(\ref{eq:nz}). However in a more realistic scenario the distribution of photometric errors,
and hence redshift distributions, will be known only up to some uncertainty.
In this section we investigate the impact of such uncertainties in the
constrainig power on growth rate by marginalizing over redshift
distributions.

For concreteness we focus in a case with only two redshift bins with $z_{mean}=\{0.78, 0.96\}$
and width $\Delta z /(1+z) = 0.1$). In our framework redshift
distributions are characterized by a width, given by $\sigma_z$ in Eq.~(\ref{eq:nz}),
and set of minimum and maximum values for the photometric top-hat
selection that determine the mean redshift $z_{mean} =
(z_{p, min}+z_{p, max})/2$. Thus, to marginalize over
miss-estimations of photometric errors, the ``width'' of $n(z)$, we
vary $\sigma_z$. To marginalize over the
``mean redshift'' of $n(z)$, we shift
both $z_{p, min}$ and $z_{p, max}$ by the same amount.
This procedure automatically changes either the width or the location of the
underlying redshift distribution. Effectively, it also marginalizes
over the amount of bin-overlap. In what follows we do not put priors on any parameter.

\begin{figure*}
\begin{center}
\includegraphics[trim = 1cm 0.4cm 1cm 1cm, width=0.5\textwidth]{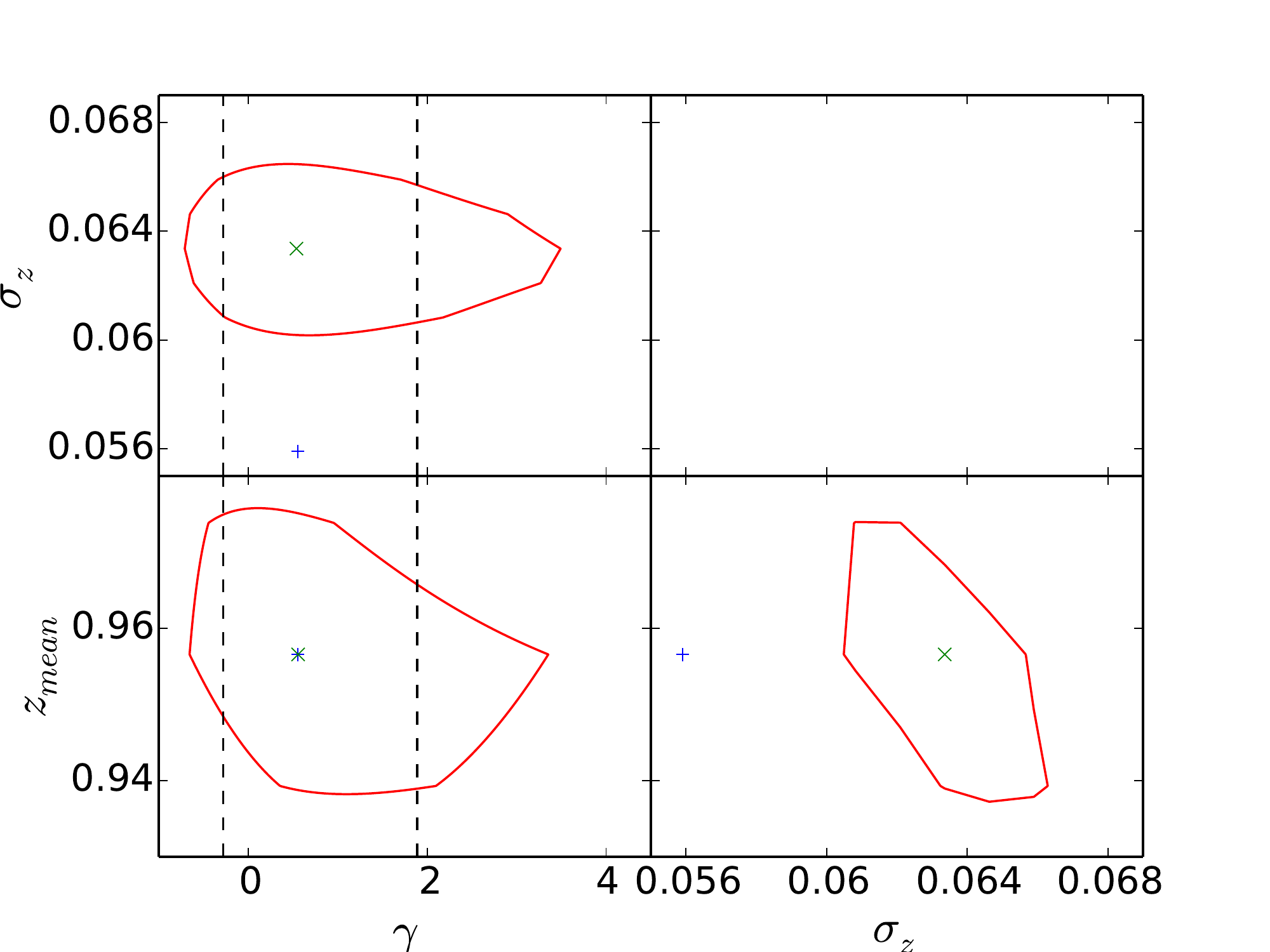}
\caption{
\emph{The impact of photo-z uncertainties
    on growth rate measurements:} We considered a
    case with only two redshift bins, with fiducial $z_{mean}=\{0.78,
    0.96\}$ for a population with $b=2$ and $\sigma_z/(1+z)=0.03$
    where the location ($z_{mean}$) and width ($\sigma_z$) of the
    redshift distribution of the second bin are free parameters (in
    addition to $\gamma$). The figure shows the 68\% confidence
    regions for $\gamma$, $z_{mean}$ and $\sigma_z$. Black dashed
    lines enclose the $1-\sigma$ region when $\gamma$ is the only free
    parameter.  Blue ``+" markers correspond to fiducial values while
    green ``x" markers correspond to the best fit value after
    marginalizing over the remaining parameter. The marginalized error
    in $\gamma$ increases by $10\%$ with respect to perfectly known
    redshift distributions (dashed lines). In turn the best-fit
     $\gamma$ is unbiased (see text for further cases).}
\label{fig:figzmin}
\end{center}
\end{figure*}

Figure~\ref{fig:figzmin} shows the $1-\sigma$ contours of the growth rate index $\gamma$, the mean
redshift of the second bin and the width of the photometric error at
this bin for the bright population with $b=2$ and
$\sigma_z/(1+z_{mean})=0.03$ (Pop 2). For this first case we have choosen to
set $z^{bin 1}_{p,max}=z_{p,min}^{bin2}$ as we marginalized over mean
redshift of bin 2. This means that we are also changing the width and
location of the redshift distribution of bin 1 (while the amount of
bin-overlap is set by the nuisance variable $\sigma_z^{bin 2}$).
From Fig.~\ref{fig:figzmin}, we find that marginalizing over $z_{mean}$
and $\sigma_z$ increases the best-fit bin width above the fiducial value
by $\%10$ but it does not bias the recovered growth rate index
(neither the mean redshift). The error on $\gamma$ increases by about $10\%$ when marginalizing over the
$z_{mean}$ and $\sigma_z$ of bin 2, compared with the case with fixed 
$z_{mean}$ and $\sigma_z$ (represented by dashed lines in Fig. \ref{fig:figzmin}). 
In turn the marginalization shows that $\sigma_z$
and $z_{mean}$ are slightly correlated (bottom right panel of Fig.~\ref{fig:figzmin}).
We performed the same marginalization for the population with $b=1$
and $\sigma_z/(1+z)=0.05$ (Pop 1), 
finding similar conclusions ($\gamma$ is recovered unbiased, with an
error $14\%$ worse).

We also considered what happens if we do not keep both bins 
sharing the same boundary in photo-z space. In this case the redshift
distribution of bin 1 is kept totally fixed through marginalization of
$n^{bin 2}(z)$
and the bin-overlap is changed by both $z_{mean}$ and $\sigma_z$ of
bin 2. In this case we find a smaller correlation between $z_{mean}$ 
and $\sigma_z$ and also a smaller marginalized error on
$z_{mean}$. The marginalized error on $\gamma$ increases by $9\%$ when considering
Pop1 and only $\%6$ for Pop2, while the best-fit value is always recovered un-biased. 

A full analysis on how to  optimize and marginalize the photo-z uncertainties
using more realistic photometric errors is beyond the scope of this
paper. 
But the results presented in this section, and Fig. \ref{fig:figzmin}, indicates that it is possible
to account for such uncertainties 
without a major loss in constraining power on growth rate measurements.

\subsubsection{The impact of shot-noise}

One strong limitation when it comes to implementing the
``multiple tracers'' technique in real spectroscopic data is the
need to have all the galaxy samples well above the shot-noise limit (at the same
time as having the largest possible bias difference), see for instance
\pcite{gil-marin2010}. This is cumbersome because spectroscopic data
is typically sampled at a rate only slightly above the shot-noise (to maximize
the area) and for pre-determined galaxy samples (e.g LRGs, CMASS).
In a photometric survey these aspects change radically because there
is no pre-selection (beyond some flux limits) and the number of sampled
galaxies is typically very large (at the expense of course of poor
redshift resolution). Therefore is interesting to investigate if the
overall density of the samples have any impact in our results. 

\begin{figure}
\begin{center}
\includegraphics[trim = 0cm 0cm 0cm 4cm, width=0.43\textwidth]{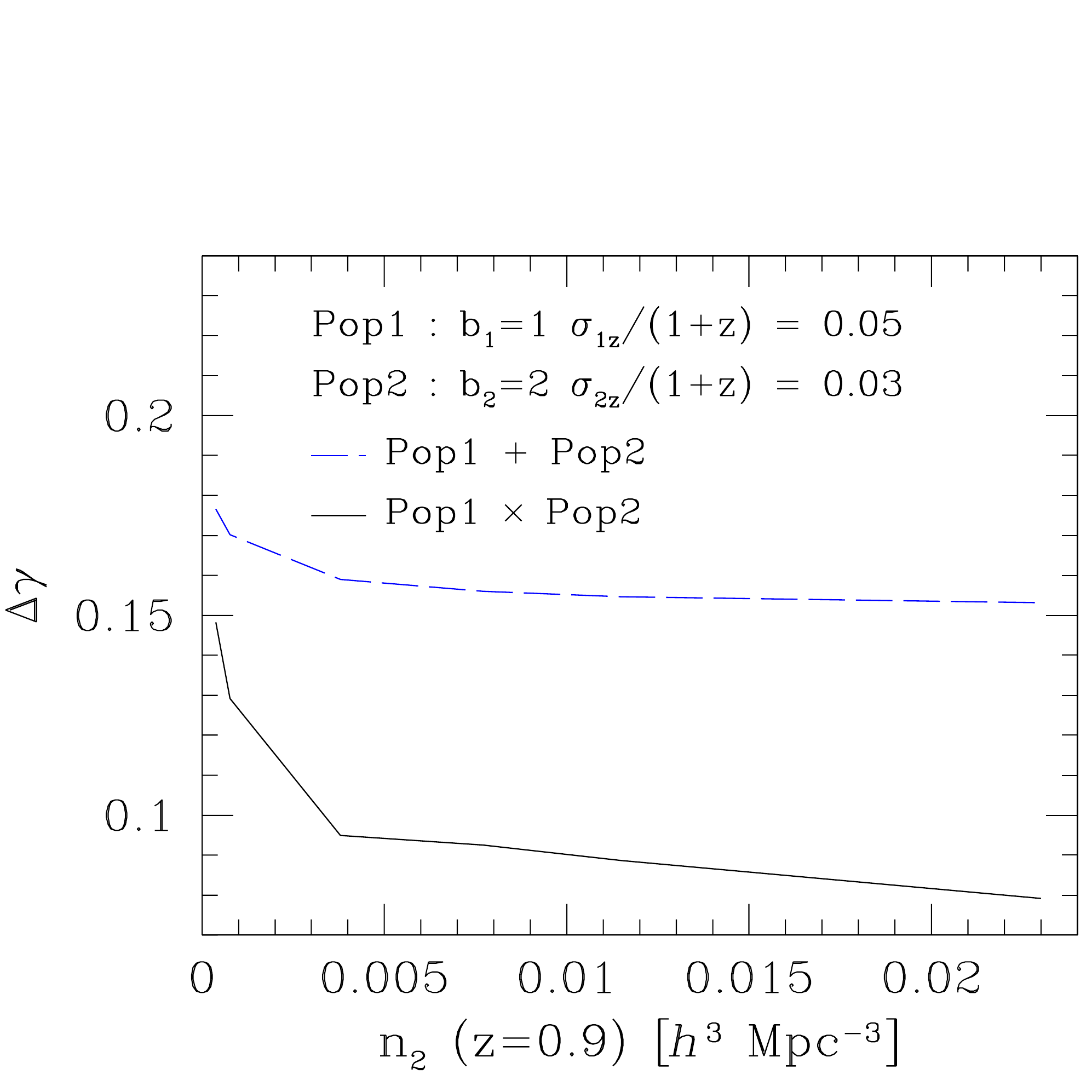}
\caption{\emph{The impact of shot noise}: We consider the combined analysis of two populations
in a redshift bin configuration of $\Delta_z/(1+z)=0.05$ and 
  show how  constrains on $\gamma$ depend on the (shot) noise level of the more biased
  population (typically the brighter, less abundant
  sample). Constrains are almost un-affected unless the density drops by an
  order of magnitude or more compared to the one of Pop1
  ($n_2=0.023\,{\it h}^{3}\,{\rm Mpc}^{-3}$).}
\label{fig:fig5}
\end{center}
\end{figure}

Figure \ref{fig:fig5} shows the constrain in $\gamma$ for the
combination of two samples, one unbiased population
with $\sigma_z/(1+z)=0.05$ and a population with $b=2$ and
$\sigma_z/(1+z)=0.03$. We keep the number density for the unbiased
population as $n(z=0.9)=1.8\times 10^{-2}\,{\it h}^3{\rm Mpc}^3$ 
while we vary the number density of the second (typically brighter)
sample\footnote{Note that we assume the same shape for $N(z)$ as given in Eq.~(\ref{eq:galaxyselectionfunction}) but we
  vary the overall normalization, which we characterize by the
  comoving number density at $z=0.9$.}. 
The solid black line corresponds to the case in
which both populations are correlated (same sky) and the dashed
blue line to different areas. 
In both scenarios we see that decreasing the number density of the second population
does not impact the error on $\gamma$ unless one degrades it by an
order of magnitude or more (below $n(z=0.9)\sim3.0\cdot 10^{-3}$).
Above this value, the error is mostly controlled by the
tracer with lower bias. 

\begin{figure}
\begin{center}
\includegraphics[width=0.45\textwidth]{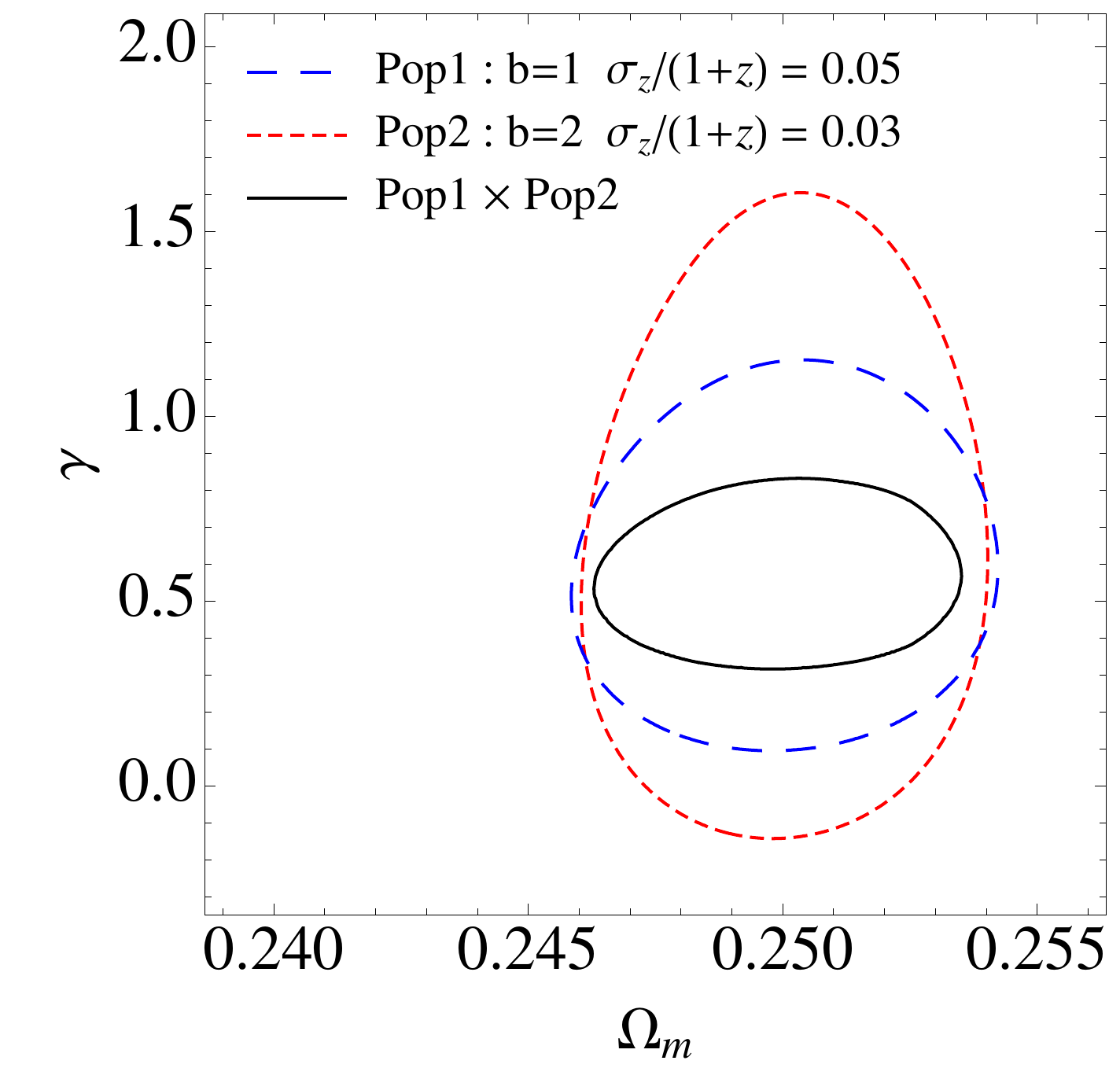}
\caption{\emph{Marginalizing over the shape of P(k) }: Contour plot of the
  posterior joint distribution when we consider both $\gamma$ and
  $\Omega_m$ as nuisance parameters. We find no significant
  degeneracies. The error on $\gamma$ degrade
  $35\%$ for Pop1, $16\%$ for Pop2 and only $6\%$ for
  Pop1$\times$Pop2. These results corresponds to a combined analysis
  of 6 bins with $\Delta z / (1+z) = 0.1$.}
\label{fig:Om}
\end{center}
\end{figure}

\begin{figure*}
\begin{center}
\includegraphics[width=0.45\textwidth]{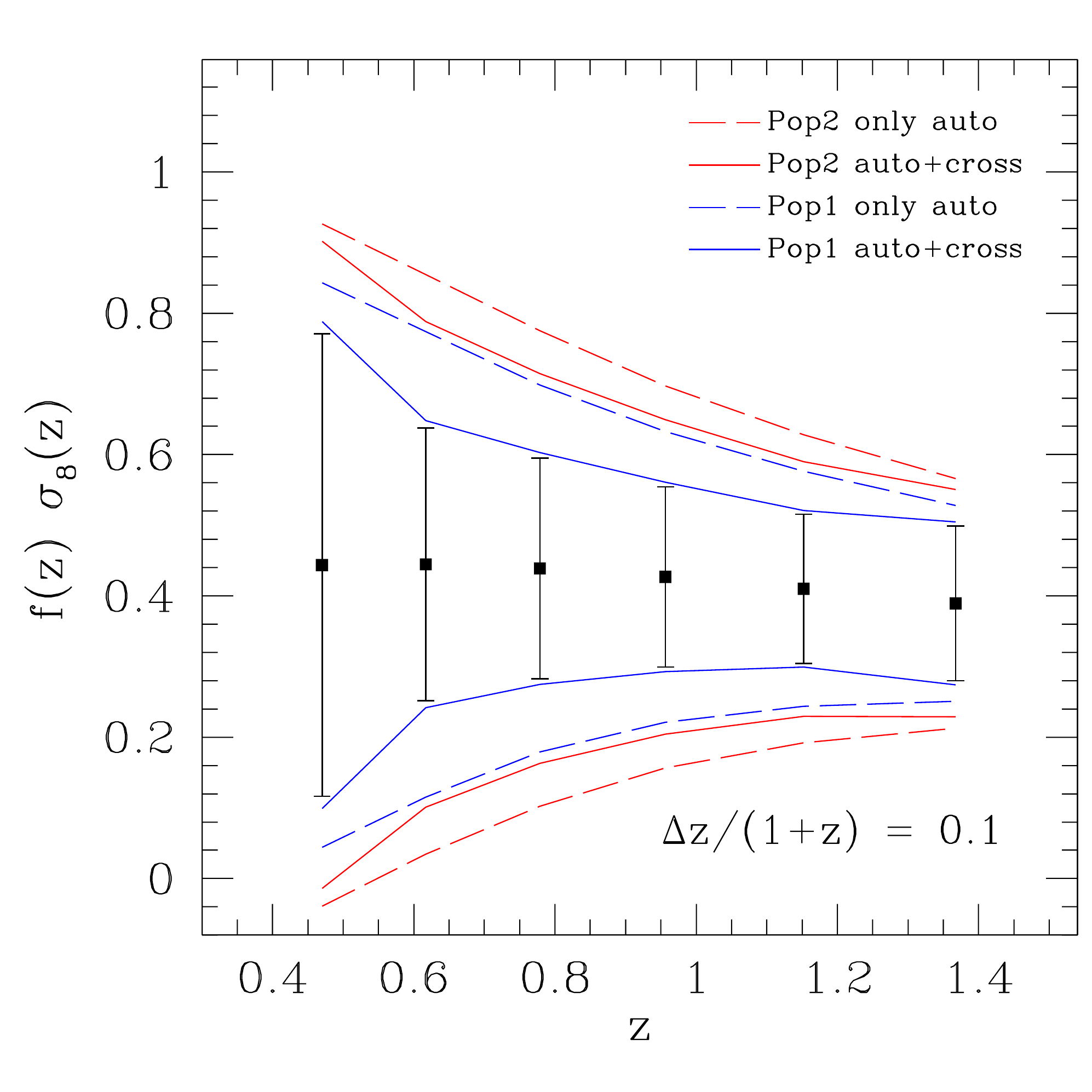}
\includegraphics[width=0.45\textwidth]{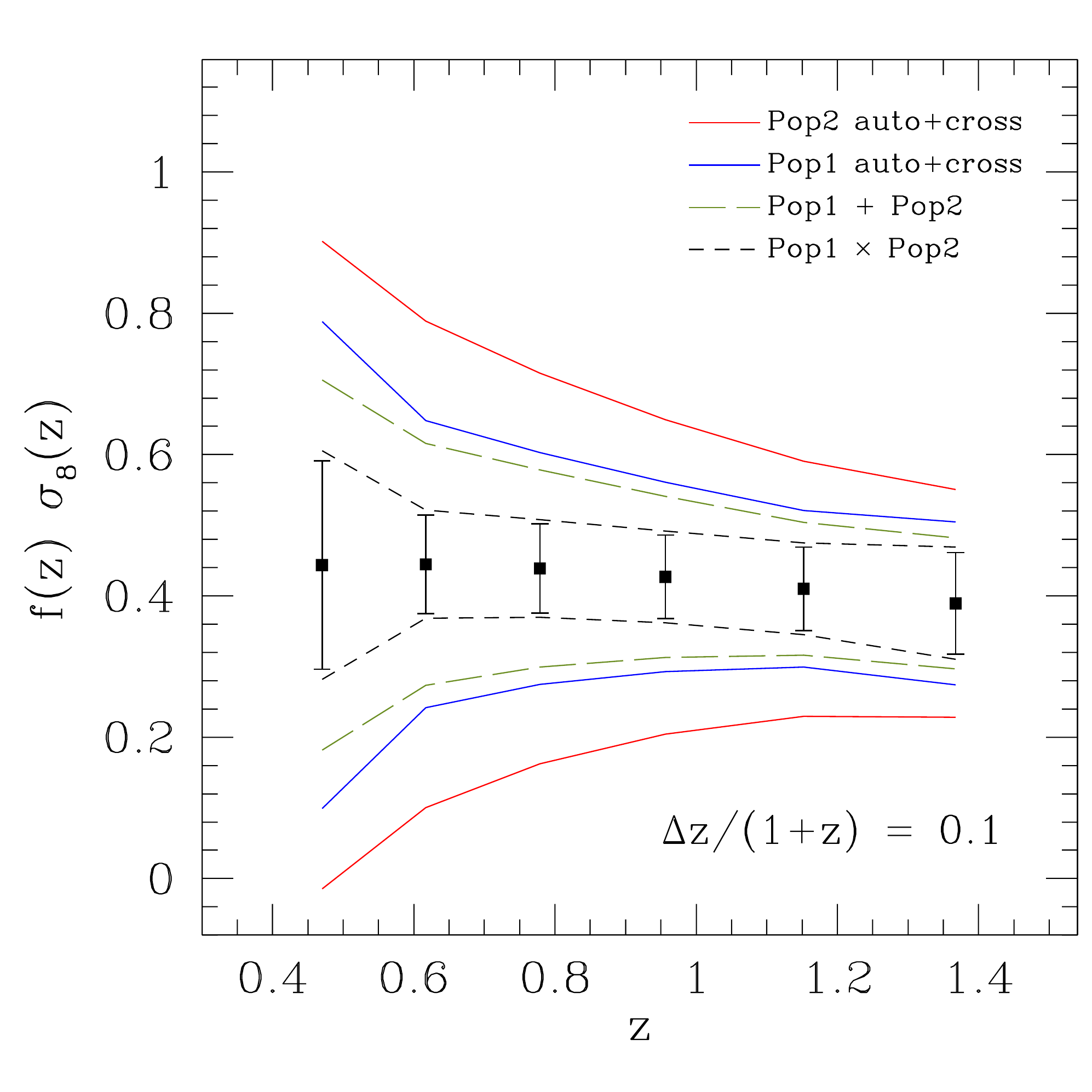}
\caption{\emph{Constrains on $f\sigma_8$}: derived at different redshift
bins, for a bin configuration of $\Delta z/(1+z)=0.1$. The left panel
focuses on one population only fits, and the gain from using
auto+cross correlations among all redshift bins as observables instead of just the auto-correlations.
The right panel stresses instead the gain from combining
two populations (through their auto and cross-correlations) either in
different patches of the sky (Pop1$+$Pop2) or the same
(Pop1$\times$Pop2). In all cases Pop1 refers to a galaxy population with
$b=1$ and $\sigma_z/(1+z)=0.05$ and Pop2 to $b=2$ and
$\sigma_z/(1+z)=0.03$. The covariance among the derived errors on
$f(z)\times \sigma_8(z)$ is taken into account in the
fit. Our results show that by using RSD with two tracers a DES-like photometric survey can place
$\sim 15\%$ constrains in the evolution of $f\sigma_8$ for several
bins in $z \gtrsim
0.8$ (with errors almost uncorrelated between bins, see text for details).}
\label{fig:fig6}
\end{center}
\end{figure*}

\subsubsection{Marginalizing over Power spectrum shape}

In the analysis presented so far, we have assumed a perfect knowledge on the shape 
of the matter power spectrum and hence of the underlying cosmological
parameters. However it is important to explore possible degeneracies
between the parameter we base on for RSD, namely $\gamma$, and other
cosmological ones. While we leave a full exploration of degeneracies
for a follow-up paper we now study the impact of varying the shape of
the power spectrum in addition to $\gamma$. We do this by considering
the matter density $\Omega_m$ as a nuisance parameter to marginalize
over. By doing so, we are mostly studying the effect of the matter power spectrum 
shape in the analysis.

For concreteness we focused on the binning configuration with $\Delta
z /(1+z) = 0.1$ (6 bins). Figure \ref{fig:Om} shows the contour plots
of the posterior distribution in the $\Omega_m-\gamma$ for Pop1
(unbiased with bad photo-z), Pop2 (biased with good photo-z)
and Pop1$ \times$ Pop2. We find that there are no significant
degeneracies and $\Omega_m$ is determined with quite good precision.

Then, if we marginalize over $\Omega_m$, we see that the errors on
$\gamma$ degrade a $16\%$ for an unbiased population, a $35\%$ for the biased one while the
effect when we cross correlate both populations is only a $6\%$ worse
error on $\gamma$. Therefore, the
conclusions obtained in previous sections are still valid, even if we allow the shape of the matter
power spectrum to change.

\subsection{Constraining the redshift evolution of the Growth Rate of Structure}

So far we have used the combined analysis of all the redshift bins to
constrain one global parameter, namely the growth rate index $\gamma$
in Eq.~(\ref{eq:fz}). We now turn into constraining $f(z)\sigma_8(z)$
itself, as a function of redshift. We use a redshift bin
configuration given by $\Delta z/(1+z)=0.1$, in the photometric
range $0.4<z<1.4$. This configuration consist of 6 bins, and
hence we fit $f(z)\sigma_8(z)$ evaluated at the mean of these 
bins. These $f\sigma_8$ values are of course correlated, and we
include the proper covariance among the measurements (i.e. we do a global fit to the 6 values simultaneously).

In the left panel of Fig.~\ref{fig:fig6} we focus on the gain from
adding cross-correlations among the bins, and show
the constrain on $f\sigma_8$ for a
single unbiased population with photometric redshift of
$\sigma_z=0.05$ (Pop 1, in blue) and also for a single tracer with bias $b=2$ and
$\sigma_z=0.03$ (Pop 2, in red). Dashed lines corresponds to using only
auto-correlations and solid to including also all the redshift bins
cross-correlations to the observables. The trend for the errors when we only use
auto-correlations are similar to the ones observed in Fig. 8 of
\pcite{2011MNRAS.415.2193R}. Although in detail we are using different widths for our
redshift bins, and we use $C_{\ell}$ while they 
used angular correlation functions, $w(\theta)$. 

As in Sec.~\ref{sec:singlepop} there is a gain
from the addition of cross-correlations, which is now split across
the bins (i.e. $20-30 \%$ for Pop1 in each of the 6 bins, and a bit less for
Pop2). 

In turn, the right panel of Fig.~\ref{fig:fig6} focuses in the gain from
combining the two tracers (and using both auto and cross correlations
among redshift bins, as in Sec.~\ref{sec:twopops}). Here the solid lines correspond to the single population
cases discussed above, while the black short-dashed line to the combined analysis
assuming these populations are correlated (same sky). For completeness, the
dashed green line is the result when these two samples are assumed independent.
Again, there is a factor of $\sim 2.5$ to be gained by combining galaxy
samples as opposed to only the unbiased sample. 

If we compare our predictions to measurements from spectroscopic surveys like VIPERS
\cite{delaTorre2013} with constrains  $f\sigma_8(z=0.8) = 0.47 \pm
0.08$ or WiggleZ \cite{blake11} where $f\sigma_8(z=0.76) = 0.38 \pm
0.04$ we find that DES can achieve the same level of errors
  ($\sim 15\%$) in determining the growth of structure but extending the constrains beyond redshift of unity. This
  is quite unique and interesting as there is, to our knowledge, no other spectroscopic survey expected
  to provide such measurements in the medium term future (before
  ESA/Euclid or DESI).

\subsubsection{Impact of unknown redshift distributions}
\label{sec:nz}

 In this section, we investigate the consequences of not having a
perfect knowledge of the redshift distributions used to project the
3D clustering into tomographic bins. For concreteness we do this by investigating how
the error on $f \sigma_8$ resulting from a single bin at $z=1.15$ change
when we also vary the assumed 
underlying redshift distribution (within the binning configuration of $\Delta z/(1+z)=0.1$). We note that in doing this we consider the full
covariance with adjacent bins while the explored parameter space
consist of 3 values of $f(z) \sigma_8(z)$ (at $z=1.05,1.15, 1.36$) and either the
mean or the width of $N(z)$ for the central bin at $z=1.15$.

We first concentrated in marginalized over the mean redshift of the assumed
$N(z)$ for the central bin assuming a flat prior of $3\%$ around
$z_{mean}=1.15$. We have repeated this for all the cases explored in
Fig.~\ref{fig:fig6}, namely we consider populations 1 and 2 individually and then
the same sky and different sky combinations of both populations.
We have not found significant changes with respect to the results in
previous sections finding differences smaller than $1\%$ for the cases with individual
populations and less than $5\%$ for the case in which we combine the two populations.

Then, we marginalize over photometric errors and we
find differences smaller than $1\%$ in the recovered constrains in
$f(z)\sigma_8(z)$ with respect to the case in which we assume perfect
knowledge of the redshift distributions. For concreteness we did this cross-check for the case Pop1$\times$Pop2 in
last two redshift bins shown in Fig.~\ref{fig:fig6}.

\subsection{The case of high-photometric accuracy}

In the previous sections, we have focused in galaxy surveys with broad-band
photometry for which the typical photometric error achieved is of the
order $0.1$ depending on galaxy sample and redshift\footnote{We assumed $0.03-0.05$ (1+z)}.
We now turn to narrow-band photometric surveys such as the ongoing PAU
or J-PAS Surveys \cite{PAU,1109.4852,paucam,Taylor2013}. 
  These surveys
are characterized by a combination of tens of narrow band (NB) filters
($\sim 100 \angstrom$) and few standard broad bands (BB) in the optical range. In
the concrete case of PAU the NB filters are $40$ in total ranging from
$\sim 4400 \angstrom$ to $\sim 8600 \angstrom$ that will perform
as a low resolution spectrograph. With the current survey
strategy, it will obtain accurate photometric redshifts for galaxies
down to $i_{AB} \sim 22.5$ for which the typical redshift
accuracy will be $\simeq 0.003 (1+z)$ (or $10 {\it h}^{-1}\,{\rm
  Mpc}$). This scenario then resembles quite
closely a purely spectroscopic survey \cite{Asorey2012}.  However 
the expected density of this sample is $\sim 10000$ galaxies
per ${\rm deg}^2$, much denser than any spectroscopic surveys to the same depth.

We do not aim here at giving a forecast for PAU but rather at
investigating the issue of combining samples with high-photometric
accuracy. Hence we will assume the same overall redshift distribution
as in Sec.~\ref{sec:survey} but consider only a total $50\times 10^6$ galaxies within $5000$
deg$^2$. This is in broad agreement with PAU specifications (see
\pcite{1109.4852} and \pcite{paucam} for further details).

We again study two populations, one corresponding to the main sample with bias $b=1$
and another to the LRG sample with $b=2$, both with a very good photometric accuracy
of $\sigma_z/(1+z)=0.003$. We consider a set of
21 narrow redshift bins of width $\Delta z = 0.003 (1+z)$ concentrated in
$0.94<z<1.06$ (hence we are only looking at a portion of the survey redshift range).

The  error on $\gamma$ 
are given in Table~\ref{table:narrowbands}, for both the new narrow-band
and the broad-band samples discussed previously. 
For a single population, 
this table shows that a factor of $\sim 10$ better $\sigma_z$ yields a factor
of $\sim 10$ gain in constraining power. The improvement in $\gamma$ seems to
increase linear with the improvement in $\sigma_z$.

\begin{table}
{\center
\begin{tabular}{ccccc}
${\rm Population}$ & b & $\sigma_z/(1+z)$ & ${\rm  Auto}$ & ${\rm Auto + Cross}$ \\ \hline
Broad Band  (BB) \\ \hline
${\rm Pop1}$ & 1 & 0.05 & 0.809  & 0.564 \\
${\rm Pop2}$ & 2 & 0.03 & 0.826  & 0.447 \\
${\rm Pop1 \times Pop2}$ & -& - & -  & 0.35 \\
${\rm Pop1+          Pop2}$ & -& - &  -  & 0.36 \\ \hline
Narrow Band  (NB) \\ \hline
${\rm Pop1}$ & 1 & 0.003 & 0.047 & 0.027 \\
${\rm Pop2}$ & 2 & 0.003 & 0.088 & 0.040 \\
${\rm Pop1 \times Pop2}$ & -& - & - & 0.016 \\
${\rm Pop1+          Pop2}$ & -& - & - & 0.023 \\ \hline 
\end{tabular}
\caption{Error in the growth rate $\gamma$ from a
  combination of $21$ narrow bins in the range $0.94<z<1.06$. 
The 4 top entries correspond to a Survey with
 Broad-Band (BB) filters: Pop1-BB assumes 
$b=1$ and $\sigma_z/(1+z)=0.05$ (``main sample'')
  while Pop2-BB has $b=2$ and $\sigma_z/(1+z) = 0.03$ (``LRG
  sample''). The 4 bottom entries correspond to 
a Survey with Narrow-Band (NB)  filters. Here Pop1 and
  Pop2 have the same bias as the BB case but much precise photo-z,
 both with $\sigma_z/(1+z)=0.003$.}
\label{table:narrowbands}
}
\end{table}

After combining the two populations, we see that the errors in
$\gamma$ for the broad-band case is similar if samples cover the
same region of sky (${\rm Pop1\times Pop2}$) or different regions 
 (${\rm Pop1+  Pop2}$).
 This is because  the redshift range
considered ($0.94<z<1.06$) is too narrow compared to $\sigma_z$ and
the cosmic variance cancelation can
not take place. Instead, for the narrow band surveys we find a
$43\%$ improvement
for the case ${\rm Pop1\times Pop2}$ with respect to ${\rm
  Pop1+Pop2}$. For the same sky case, the final
error is  $\Delta \gamma \simeq 0.0163\times(5000\,{\rm deg}^2/{\rm Area})^{1/2}$, in such a way
that even a moderate survey of $250\,{\rm deg}^2$ could achieve 
$\Delta \gamma \sim 0.07$. In that same narrow redshift range, DES
yields an error 5 times worse with 20 times better area (but note that
in the case of small areas we could be limited by the $\ell_{min}$, the
largest scales available).

\section{Conclusions}\label{sec:conclusions}

We have studied how measurement of redshift-space distortions (RSD) in
wide field photometric surveys produce constrains on the growth of structure, in the linear
regime.  We focused in survey specifications similar to those of
  the ongoing DES or PanSTARRS, that is, covering about $1/8$ of sky up
  to $z \sim 1.4$, and targeting galaxy samples with photometric
  redshift accuracies of $0.03-0.05 (1+z)$ (and hundred of million
  galaxies prior to sample selection). We also show results for
ongoing photometric surveys, such as PAU and J-PAS, that have a much
better photometric accuracy.

First, we have found that for a single population 
we can reduce the errors in half by including all the
cross-correlations between radial shells in the analysis. This is
because one includes large scale radial information that was missed when only
considering the auto-correlations of each bin. The final constraining
power depends on the details of the population under consideration,
in particular the bias and the  photometric accuracy. Less bias
gives more relative importance to RSD in the clustering amplitudes. In
turn, better photo-z allows for narrower binning in the analysis and
more radial information.  We find that the $\gamma$ constrains depend
roughly linearly in both bias or $\sigma_z$. This means that for 
10 times better photo-z errors, such as in PAU, we can improve by 10
the cosmological constrains.

Typically less bias implies a fainter sample,
with worse photo-z, therefore these quantities compete in determining
the optimal sample. Furthermore we find that optimal constrains are achieved
for bin configurations such that $\Delta z \sim \sigma_z$. Although
the optimal errors depend on the details of the galaxy sample and binning
strategy, the gains from adding cross-correlations are very robust in
front of these variations.

In order to avoid sample variance, we have also considered 
what happens if we combine the measurement
of RSD using two different tracers. 
This is motivated by the idea put forward in \pcite{McDonaldSeljak}
  for the case of spectroscopic (hence 3D) redshift surveys, where the
  over-sampling of (radial + transverse) modes allows a much better 
  precision in growth rate constrains, as long as samples are in the low shot-noise limit.
Combining auto and cross angular correlations in redshift bins,
we find that if we assume that both tracers are independent,
which corresponds to samples from different regions on the sky, the
constrains on the growth of structure parameters improve a 30-50\%
(due to the fact that one has doubled the area). Remarkably if we consider that the populations are not independent, i.e., they trace
the same field region, we find an overall improvement of $\sim 2-3$ with
respect to single populations when constraining $\gamma$. This means
that there is a large potential gain when sampling the same modes more
than once.

Translating into actual constrains this implies that a DES-like
photometric survey should be able to measure the growth rate of structure $\gamma$ to
an accuracy of $5-10\%$ from the combination of two populations and all
the auto+cross correlations in the range $0.4<z<1.4$ (see
Fig.~\ref{fig:fig1}). Even though these values correspond to a survey
of $5000 {\rm deg}^2$ ($f_{sky}=0.125$) they should scale as $f_{sky}^{-1/2}$ for a
different area, given our assumptions for the covariance in Eq.~(\ref{eq:CovObs2D}).

In Fig.~\ref{fig:fig5}, we have shown that constrains weaken
once one of the populations enter a shot-noise dominated regime, as is
typical of spectroscopic samples. However 
one needs to dilute over 10 times the number densities
for a photometric survey, such as DES, for this to happen.
Thus, as shown in Section 3.4, by improving on photo-z accuracy 
without much lost of completeness,
a photometric sample can in fact outperform a diluted spectroscopic
version with similar depth and area (see also \pcite{1109.4852}).
\begin{figure}
\begin{center}
\includegraphics[width=0.47\textwidth]{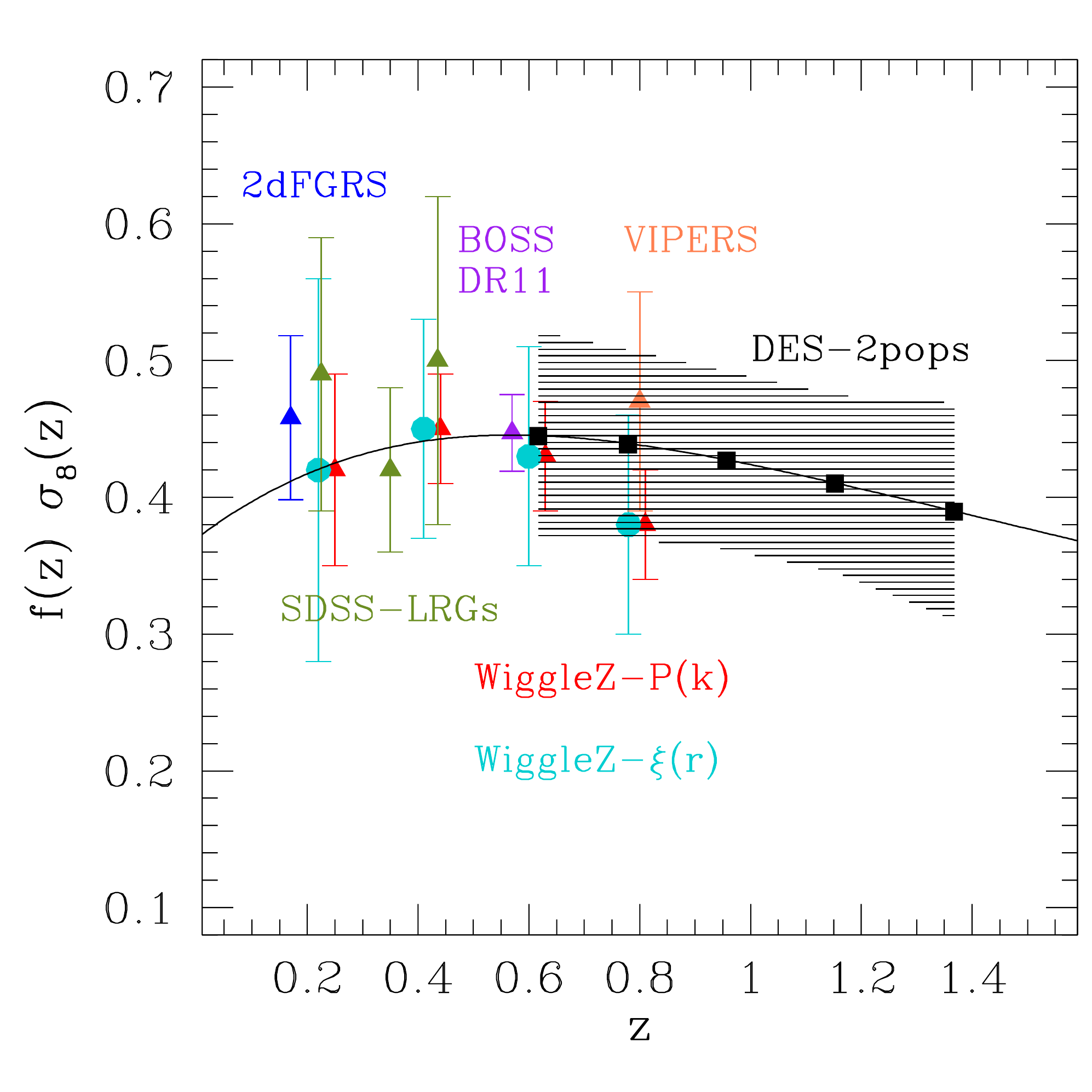}
\caption{Combined constrains in the evolution of the growth rate
  of structure from spectroscopic data, 2dFGRS, SDSS-LRGs, Wiggle-Z,  VIPERS
  and BOSS (see text for details), and forecasted for DES using two
  photometric populations (same as in Fig. 8). The addition of 
  DES (shaded area) allows to trace the growth rate of structure all the way to $z \sim 1.4$.}
\label{fig:fig9}
\end{center}
\end{figure}
In this paper we focused on large angular scales where the
approximation of linear and deterministic bias and linear RSD should
hold (see for instance \pcite{2011MNRAS.414..329C}). 
Although we set $\ell_{max}\sim 200$, much of the constraining power in our results,
given the typical size of our redshift bins, comes from larger
scales, $\ell \lesssim 40$. Yet, a more realistic assessment of these
aspects will need to resort to numerical simulations. We leave this for future work.

Lastly, we also investigated what constrains can be placed with this
method in the evolution of the growth rate of structure,
$f(z)\times\sigma_8(z)$. 
We found that binning two DES populations
into $6$ bins in the range $0.4<z<1.4$ yields constrains on
$f(z)\times\sigma_8(z)$ of $\sim 15\%$ for each bin above $z \sim
0.6$. This is shown in Fig.~\ref{fig:fig9}.
That case corresponded to bin widths larger than the photometric errors
of the samples, which may not be optimal but yield constrains
almost uncorrelated between bins
($\rho_{ij}\sim -0.05$)\footnote{For bins ($i,j$) we define the cross-correlation
  coefficient $\rho_{ij}$ as 
  $\rho_{ij}={\rm Cov}_{ij}/\sqrt{{\rm Cov}_{ii}{\rm Cov}_{jj}}$ with
  ${\rm Cov}_{ij} = \langle (x - \langle x \rangle)_i (x -
  \langle x \rangle)_j \rangle$ and where $x$ stands for $f \times \sigma_8$.
}. A narrower binning, $\Delta z/(1+z)=0.05$ leads to better
constrains per bin, $\Delta(f\sigma_8) \sim 10\%$, at the
expense of more correlation between bins, $0.2< \rho_{ij}<0.65$.

In addition to the DES forecast (shadowed region) we over-plot in Fig. \ref{fig:fig9} current
constrain from spectroscopic surveys, 2dFGRS \cite{2df}, LRG's
from SDSS (\pcite{tegmark06} and \pcite{cabre09}), WiggleZ either from power spectrum
\cite{blake11} or correlation function \cite{contreras13}, and the recent VIPERS, \cite{delaTorre2013}  and BOSS results
\cite{samushia14}. Note that these values are not expected to improve radically
in the near future. This implies that DES will be able to add quite
competitive constrains 
in a redshift regime unexplored otherwise with spectroscopic surveys (i.e. $z \gtrsim 1$),
yielding a valuable redshift leverage for understanding the nature of
dark energy and cosmic acceleration through the growth of structure.

\section*{Acknowledgments}
We thank Pablo Fosalba, Will Percival and Ashley Ross for comments on the draft paper
and helpful discussions.
Funding for this project was partially provided by the
Spanish Ministerio de Ciencia e Innovacion (MICINN),
project AYA2009-13936, AYA2012-39559, Consolider-Ingenio CSD2007- 00060,
European Commission Marie Curie Initial
Training Network CosmoComp (PITN-GA-2009-238356),
research project 2009-SGR-1398 from Generalitat de Catalunya
and  the Ramon y Cajal MEC program.
J. A. was supported by the JAE program grant from the Spanish National Science
Council (CSIC) and by the Department of Energy and the University of Illinois at Urbana-Champaign.

\bsp

\label{lastpage}


\begin{thebibliography}{99}



\bibitem[{Asorey et al.}~<2012>]{Asorey2012} 
Asorey J., Crocce M, Gazta{\~n}aga E., Lewis A., 2012, MNRAS  {\bf 427}, 1891

\bibitem[{Banerji et al. }~<2008>]{Banerji2008} 
Banerji M., Abdalla F. B., Lahav O., Lin, H., 2008, MNRAS  {\bf 386}, 1219

\bibitem[{Benitez et al.}~<2009>]{PAU}
Benitez N., et al., 2009, ApJ {\bf 691}, 241

\bibitem[{Benjamin et al.}~<2013>]{benjamin2013}
Benitez N., et al., 2013, MNRAS {\bf 431}, 1547


\bibitem[{Blake et al.}~<2007>]{blake07}
Blake C., Collister A., Bridle S., Lahav O., 2007, MNRAS {\bf 374}, 1527

\bibitem[{Blake} et al.~<2011>]{blake11}
Blake C., et al., 2011 ,MNRAS {\bf 415}, 2876B

\bibitem[{Bonvin C., Durrer R.}<2011>]{PhysRevD84063505}
Bonvin C., Durrer R., 2011, Phys. Rev. D {\bf 84}, 063505

\bibitem[{Cabr{\'e} et al.}<2007>]{cabre07} 
Cabr{\'e} A., Fosalba P., Gazta{\~n}aga E., Manera M., 2007, MNRAS {\bf 381}, 1347

\bibitem[{Cabr{\'e} \& Gazta{\~n}aga}~<2009>]{cabre09}
Cabr{\'e} A., Gazta{\~n}aga E., 2009, MNRAS {\bf 393} ,1183

\bibitem[{Cai \& Bernstein} <2012>]{1112.4478} 
Cai Y. C., Bernstein G., 2012, MNRAS {\bf 422}, 1045

\bibitem[{Castander et al.}~<2012>]{paucam}
Castander F. J., et al., 2012, Proceedings of the SPIE {\bf 8446}, 84466D

\bibitem[{Challinor \& Lewis}<2011>]{2011arXiv1105.5292C}
Challinor A., Lewis A., 2011, Phys. Rev. D {\bf 84}, 043516

\bibitem[{Crocce, Cabr{\'e}, \& Gazta{\~n}aga}<2011>]{2011MNRAS.414..329C}
Crocce M., Cabr{\'e} A., Gazta{\~n}aga E., 2011, MNRAS {\bf 414}, 329

\bibitem[{Contreras et al.}~<2013>]{contreras13}
Contreras C, et al., 2013, MNRAS {\bf 430}, 924

\bibitem[{Crocce et al.}<2011>]{2011arXiv1104.5236C}
Crocce M., Gazta{\~n}aga E., Cabr{\'e} A., Carnero A., S{\'a}nchez E.,
2011, MNRAS {\bf 417}, 2577



\bibitem[{de la Torre et al.}<2013>]{delaTorre2013}
de la Torre S., Guzzo L., Peacock J. A., et al., 2013, A\&A submitted,
e-print arXiv:1303.2622

\bibitem[{de Putter R., Dor\'e O. \& Takada M.}<2013>]{dePutter2013}
de Putter R., Dor\'e O. \& Takada M., 2013, [eprint arXiv::1308.6070]




\bibitem[{Fisher, Scharf \& Lahav} <1994>]{fisher94}
Fisher K. B., Scharf C. A., Lahav O., 1994, MNRAS {\bf 266}, 219


\bibitem[{Gazta{\~n}aga et  al.} <2012>]{1109.4852} 
Gazta{\~n}aga E., Eriksen M., Crocce M., Castander F. J., Fosalba P.,
Marti P., Miquel R., Cabr{\'e} A., 2012, MNRAS {\bf 422}, 2904

\bibitem[{Gil-Mar\' in et  al.} <2010>]{gil-marin2010} 
Gil-Mar{\' i}n H., Wagner C., Verde L., Jimenez R., Heavens A. F.,
2010, MNRAS {\bf 407}, 772

\bibitem[{Guzzo} et al.~ <2009>]{guzzo09}
Guzzo L., et al., 2009, Nature {\bf 451} ,451 

\bibitem[{Hamilton} <1998>]{hamilton98}
Hamilton A. J. S., 1998,``Linear redshift distortions: A review'', in ``The
Evolving Universe'', ed. D. Hamilton, pp. 185-275 (Kluwer Academic,
1998) [eprint arXiv: astro-ph/9708102]

\bibitem[{Kaiser}<1987>]{Kai}
Kaiser N., 1987, MNRAS, {\bf 227}, 1

\bibitem[{Kazin} et al.~<2013>]{kazin13} 
Kazin E. A., et al., 2013, MNRAS, {\bf 435}, 64

\bibitem[{Kirk et al.}<2013>]{kirk2013}
Kirk D., Lahav O., Bridle S., Jouvel S., Abdalla F., Frieman J., 2013, [eprint arXiv:1307.8062]



\bibitem[{Lewis et al.} <2000>]{cambt} 
Lewis A., Challinor A., Lasenby A., 2000, ApJ {\bf 538}, 473

\bibitem[{Lewis \& Challinor.} <2007>]{lewis2007} 
Lewis A., Challinor  A.  2007 Phys. Rev. D {\bf76}, 083005

\bibitem[{Limber}<1954>]{1954ApJ...119..655L}
Limber  D. N., 1954, ApJ {\bf 119}, 655

\bibitem[{Linder}<2005>]{linder1}
Linder, E. V. 2005, Phys. Rev. D {\bf 72}, 043529

\bibitem[{LoVerde \& Afshordi}~<2008>]{LoVerde2008}
LoVerde M., Afshordi N., Phys. Rev. D, {\bf 78}, 123506 (2008)

\bibitem[{McDonald \& Seljak}~<2009>]{McDonaldSeljak}
McDonald P., Seljak U., 2009, JCAP {\bf 0901}, 007

\bibitem[{Montanari \& Durrer}~<2012>]{1206.3545}
Montanari F., Durrer R., 2012, Phys. Rev. D {\bf 86}, 063503

\bibitem[{Newman}~<2008>]{newman08}
Newman J. A., 2008, ApJ {\bf 684}, 88

\bibitem[{Nock, Percival,\& Ross}<2010>]{2010MNRAS.407..520N} 
Nock K., Percival W. J., Ross A. J., 2010, MNRAS {\bf 407}, 520

\bibitem[{Okumura} et al.~<2008>]{okumura08}
Okumura T., Matsubara T., Eisenstein D. J., Kayo I., Hikage C., 
Szalay A. S., Schneider D.P., 2008, ApJ {\bf 676}, 889 

\bibitem[{Padmanabhan et al.} <2007>]{nikhil}
Padmanabhan N., et al., 2007, MNRAS {\bf 378}, 852

\bibitem[{Percival et al.}~<2004>]{2df}
Percival W., et al., 2004, MNRAS {\bf 353}, 1201


\bibitem[{Reid} et al.~<2012>]{reid12}
Reid B., et al., 2012, MNRAS {\bf 426}, 2719

\bibitem[{Ross et al.}<2011>]{2011MNRAS.415.2193R}
Ross A.~J., Percival W.~J., Crocce M., Cabr{\'e} A., Gazta{\~n}aga E.,
2011, MNRAS  {\bf 415}, 2193

\bibitem[{Samushia} et al.~<2014>]{samushia14}
Samushia L., et al., 2014, MNRAS {\bf 439}, 3504



\bibitem[{Taylor} et al.~<2013>]{Taylor2013}
Taylor K., et al., 2013, arXiv:1301.4175 

\bibitem[{Tegmark} et al.~<2006>]{tegmark06}
Tegmark M. et al., 2006, Phys. Rev D {\bf 74}, 123507

\bibitem[{Abbott} et al.~<2005>]{deswhitepaper}
The Dark Energy Survey Collaboration, White Paper submitted to the Dark Energy Task Force, [eprint arXiv::astro-ph/0510346]

\bibitem[{Thomas, Abdalla \& Lahav}<2011>]{thomas2011}
Thomas, S. A., Abdalla, F. B., Lahav, O. 2011,  MNRAS {\bf 412}, 1669

\bibitem[{White, Song \& Percival}~<2009>]{white2009}
White M., Song Y-S., Percival W. J., 2009, MNRAS, {\bf 397}, 1348

\end{thebibliography}
\end{document}